\documentclass[aps,prd,superscriptaddress,showkeys,nofootinbib]{revtex4-2}
\usepackage[english]{babel}
\usepackage{amsmath,bm}
\usepackage{amssymb}
\usepackage{enumitem}
\usepackage{textcomp}
\usepackage{graphicx}
\usepackage{dsfont}
\usepackage{color}

\usepackage[breaklinks=true]{hyperref}
\hypersetup{
	colorlinks=true,
	linkcolor=blue,
	filecolor=magenta,      
	urlcolor=blue,
	citecolor=red,
}

\usepackage{mathrsfs}
\usepackage[normalem]{ulem}

\usepackage{placeins}

\begin{document}

{\hfill CERN-TH-2021-185}
	
	\title{Hypermagnetogenesis from axion inflation: Model-independent estimates}
	
	\author{E.V.~Gorbar}
	\affiliation{Physics Faculty, Taras Shevchenko National University of Kyiv, 64/13, Volodymyrska Street, 01601 Kyiv, Ukraine}
	\affiliation{Bogolyubov Institute for Theoretical Physics, 14-b, Metrologichna Street, 03143 Kyiv, Ukraine}
	
	\author{K.~Schmitz}
	\email{kai.schmitz@cern.ch}
	\affiliation{Theoretical Physics Department, CERN, 1211 Geneva 23, Switzerland}
	
	\author{O.O.~Sobol}
	\email{oleksandr.sobol@epfl.ch}
	\affiliation{Institute of Physics, Laboratory of Particle Physics and Cosmology, \'{E}cole Polytechnique F\'{e}d\'{e}rale de Lausanne, CH-1015 Lausanne, Switzerland}
	\affiliation{Physics Faculty, Taras Shevchenko National University of Kyiv, 64/13, Volodymyrska Street, 01601 Kyiv, Ukraine}
	
	\author{S.I.~Vilchinskii}
	\affiliation{Physics Faculty, Taras Shevchenko National University of Kyiv, 64/13, Volodymyrska Street, 01601 Kyiv, Ukraine}
	\affiliation{D\'{e}partement de Physique Th\'{e}orique, Center for Astroparticle Physics, Universit\'{e} de Gen\`{e}ve, 1211 Gen\`{e}ve 4,  Switzerland}
	
	\date{\today}
	
	\begin{abstract}
	Axion inflation coupled to the Standard Model (SM) hypercharge gauge sector represents an attractive scenario for the generation of primordial hypermagnetic fields. The description of this scenario is, however, complicated by the Schwinger effect, which gives rise to highly nonlinear dynamics. Hypermagnetogenesis during axion inflation in the absence of nonlinear effects is well studied and known to result in a hypermagnetic energy density that scales like $H^4\,e^{2\pi\xi}/\xi^5$, where $\xi$ is proportional to the time derivative of the axion-vector coupling in units of the Hubble rate $H$. In this paper, we generalize this result to the full SM case by consistently taking into account the Schwinger pair production of all SM fermions. To this end, we employ the novel gradient-expansion formalism that we recently developed in [\href{https://arxiv.org/abs/2109.01651}{arXiv:2109.01651}], and which is based on a set of vacuum expectation values for bilinear hyperelectromagnetic functions in position space. We parametrize the numerical output of our formalism in terms of three parameters ($\xi$, $H$, and $\Delta$, where the latter accounts for the damping of subhorizon gauge-field modes because of the finite conductivity of the medium) and work out semianalytical fit functions that describe our numerical results with high accuracy. Finally, we validate our results by comparing them to existing estimates in the literature as well as to the explicit numerical results in a specific inflationary model, which leads to good overall agreement. We conclude that the systematic uncertainties in the description of hypermagnetogenesis during axion inflation, which previously spanned up to several orders of magnitude, are now reduced to typically less than 1 order of magnitude, which paves the way for further phenomenological studies.

	\end{abstract}
	
	\maketitle
	
\section{Introduction}
\label{sec-intro}

Baryonic matter in the Universe mostly exists in the form of plasma. Being composed of free-streaming charged particles, plasma very efficiently screens electric fields. On the other hand, its large electric conductivity keeps magnetic fields frozen for a long time. Therefore, it is not surprising that magnetic fields are observed everywhere in the Universe, namely, in stars, galaxies, and clusters of galaxies \cite{Grasso:2001,Kronberg:1994,Widrow:2002,Giovannini:2004,Kandus:2011,Vallee:2011,Ryu:2012,Durrer:2013,Subramanian:2016}. There exist several astrophysical mechanisms that could be responsible for the generation of magnetic fields on these length scales. In contrast, the evidence for magnetic fields in voids with coherent lengths of the order of megaparsecs based on the observation of blazars~\cite{Tavecchio:2010,Ando:2010,Neronov:2010,Tavecchio:2011,Dolag:2010,Dermer:2011,Taylor:2011,Huan:2011,Vovk:2012,Caprini:2015,Batista:2021} is quite unexpected and fascinating. Indeed, the small matter content in voids makes the direct generation of magnetic fields in voids impossible. Although these fields could be induced by outflows of magnetized matter from galaxies \cite{Furlanetto:2001,Bertone:2006,Samui:2018,Garcia:2020}, such outflows would need to be strong and coherent over tens of millions of years, which appears implausible. As a consequence, a cosmological origin of these magnetic fields emerges as an interesting and realistic possibility.

Inflationary magnetogenesis \cite{Turner:1988,Ratra:1992} naturally addresses the large coherence length of magnetic fields observed in voids. In addition, cosmological magnetic fields generated in the early Universe provide the necessary seeds for magnetic fields in protogalaxies, whose amplification through adiabatic compression \cite{Grasso:2001} and different types of dynamo processes \cite{Zeldovich:1980book,Lesch:1995,Kulsrud:1997,Colgate:2001,Vazza:2018} could easily explain the magnetic fields observed in galaxies and clusters of galaxies today \cite{Parker:1971}. One of the most attractive features of inflation for cosmology is that it results in an isotropic and homogeneous Universe consistent with the smallness of the temperature fluctuations in the cosmic microwave background. Since magnetic fields are not enhanced in a conformally flat inflationary background \cite{Parker:1968}, this means that the conformal symmetry of Maxwell's equations should be broken to ensure the possibility of inflationary magnetogenesis. Although this breaking can be done in many ways (see, e.g., Refs.\cite{Turner:1988,Ratra:1992,Garretson:1992,Dolgov:1993} or Ref.~\cite{Giovannini:2021dso} for a recent effective field-theoretical analysis), we consider in this work the axial coupling of the Standard Model (SM) hyperelectromagnetic field to a pseudoscalar ``axion'' inflaton field \cite{Anber:2006,Anber:2010,Durrer:2011,Barnaby:2012,Caprini:2014,Anber:2015,Ng:2015,Cheng:2015,Fujita:2015,Adshead:2015,Adshead:2016,Notari:2016,Domcke:2018,Cuissa:2018,Shtanov:2019,Shtanov:2019b,Sobol:2019,Domcke:2019bar,Domcke:2019,Domcke:2020,Kamarpour:2021a,Kamarpour:2021b,Ballardini:2019rqh,Giovannini:2021xbi,Giovannini:2021dso,Giovannini:2021thf,Giovannini:2021due,Caravano:2021bfn,Tripathy:2021sfb}. This scenario results in helical hypermagnetic fields, which enhances the chance of their survival in the primordial plasma.

The production of hypercharge gauge fields during axion inflation is subject to several nonlinear effects, which highly complicates the theoretical analysis. These effects include (i)~the backreaction of the produced gauge fields on the evolution of the inflation field~\cite{Cheng:2015,Notari:2016,Domcke:2020,DallAgata:2019} as well as (ii)~the Schwinger pair production of hypercharged matter degrees of freedom~\cite{Sauter:1931,Heisenberg:1936,Schwinger:1951}. The produced pairs of particles and antiparticles quickly form an ultrarelativistic plasma, which efficiently screens the electric field. This strongly affects the generation and evolution of electromagnetic fields, especially near the end of inflation and during reheating \cite{Domcke:2018,Sobol:2019,Domcke:2019,Kobayashi:2014,Froeb:2014,Bavarsad:2016,Stahl:2016a,Stahl:2016b,Hayashinaka:2016a,Hayashinaka:2016b,Sharma:2017,Bavarsad:2018,Geng:2018,Hayashinaka:2018,Hayashinaka:thesis,Giovannini:2018a,Banyeres:2018,Stahl:2018,Kitamoto:2018,Sobol:2018,Shtanov:2020,Tangarife:2017,Chua:2019,Shakeri:2019,Gorbar:2019,Sobol:2020Sch}. A quantitative description of hypermagnetogenesis during axion inflation is therefore theoretically challenging, which is why up to now only some order-of-magnitude estimates of its efficiency have been worked out in the literature~\cite{Domcke:2018,Domcke:2019bar}.

The goal of this paper is to improve on this situation, leveraging the quantitative accuracy of the gradient-expansion formalism~\cite{Sobol:2019} that we recently successfully applied to axion inflation coupled to the SM hypercharge gauge field in Ref.~\cite{Gorbar:2021}. As demonstrated in Ref.~\cite{Gorbar:2021}, this novel gradient-expansion formalism allows us to consistently account for the above-mentioned nonlinear effects, which enables us to evaluate the efficiency of gauge-field and fermion production during axion inflation at unprecedented accuracy. The basic idea behind the formalism is to consider vacuum expectation values of a truncated set of bilinear electromagnetic functions in coordinate space rather than momentum space. Solving the equations underlying our formalism in a single numerical run, we are able to describe the evolution of the electric and magnetic energy densities at percent-level accuracy during the whole inflation stage without the need for an iterative procedure. The formalism also takes into account the fact that the number of relevant gauge-field modes constantly grows during inflation as new modes become tachyonically unstable by adding appropriate boundary terms in the equations of motion for the bilinear electromagnetic functions.

In Ref.~\cite{Gorbar:2021}, we provided a detailed description of the gradient-expansion  formalism and confirmed its validity by comparing its numerical output to existing results in the literature for specific model and parameter benchmark scenarios. In this paper, we shall continue our investigation of hypermagnetogenesis during axion inflation based on the gradient-expansion formalism, now turning to a model-independent analysis. In the following, we will study the efficiency of gauge-field production during axion inflation in terms of a minimal number of parameters (the gauge-field production parameter $\xi$, Hubble rate $H$, and damping factor $\Delta$; see Sec.~\ref{sec-model} for the precise definition of these quantities), which will provide us with numerical results that are applicable across a large range of models based on different types of scalar potentials. In fact, as we will show, our numerical results will always provide a good estimate of the efficiency of gauge-field production whenever the three parameters $\xi$, $H$, and $\Delta$ vary only very slowly during axion inflation, such that their time dependence can be approximately neglected. To facilitate the application of our numerical results in future studies, we will also present semianalytical fit functions that reproduce our full numerical results to very good accuracy. These fit functions are compact and ready to use, which means that, in future studies, it will not be necessary to implement our full gradient-expansion formalism and redo the entire numerical analysis.

Finally, we will also compare the model-independent estimates in Refs.~\cite{Domcke:2018,Domcke:2019bar} to our new model-independent estimates and present fit functions for the estimates in Refs.~\cite{Domcke:2018,Domcke:2019bar}. An important outcome of this exercise will be that, while the estimates in Refs.~\cite{Domcke:2018,Domcke:2019bar} span several orders of magnitude, our new estimates are capable of reducing the uncertainty in the description of hypermagnetogenesis (without specifying a concrete model and solving the equations of the gradient-expansion formalism explicitly) down to roughly less than 1 order of magnitude. This becomes particularly apparent when comparing the explicit outcome of a specific inflationary model to three available model-independent estimates (i.e., the two estimates in Refs.~\cite{Domcke:2018,Domcke:2019bar} as well as our new estimate). 

The paper is organized as follows. In the next section, we will review the gradient-expansion formalism that was first developed in Ref.~\cite{Sobol:2019} and then further refined in Ref.~\cite{Gorbar:2021}. In particular, we will slightly adapt our notation compared to our earlier work so as to account for the fact that we are now dealing with constant values of the three parameters $\xi$, $H$, and $\Delta$. In Sec.~\ref{sec-numerical}, we will then represent the numerical output of our formalism after scanning over the three-dimensional parameter space of our model. We will specifically construct fit functions for our results as well as for the estimates in Refs.~\cite{Domcke:2018,Domcke:2019bar} and validate our approach by comparing it to the explicit results in a specific model: a simple $m^2\phi^2/2$ model for three different values of the axion--vector coupling constant. The good agreement between our model-independent estimates and the explicit numerical results in this model implies that the results presented in this paper provide a good description of all scenarios of axion-driven hypermagnetogenesis that close to the origin in field space are characterized by a simple quadratic mass term. Finally, we will summarize our findings and conclude in Sec.~\ref{sec-concl}. In the Appendix, we collect a number of numerical fit coefficients that enter the constructions of our fit functions. Throughout the paper, we use natural units and set $\hbar=c=1$ with the reduced Planck mass equal to $M_{\mathrm{P}}=(8\pi G)^{-1/2}=2.435\times 10^{18}\,$GeV. We assume that the Universe is described by a spatially flat Friedmann-Lema\^{i}tre-Robertson-Walker metric in terms of cosmic time, $g_{\mu\nu}={\rm diag\,}\{1, \,-a^{2}(t),\, -a^{2}(t),\, -a^{2}(t)\}$.

\section{Gradient expansion formalism}
\label{sec-model}

Let us consider the Abelian gauge field $A_{\mu}$ (which we will identify with the SM hypercharge gauge field shortly) axially coupled to the pseudoscalar axion inflaton field $\phi$. The corresponding action has the form
\begin{equation}
	\label{S}
	S_{A}=\int d^{4}x\sqrt{-g} \left[-\frac{1}{4}F_{\mu\nu}F^{\mu\nu}-\frac{1}{4}I(\phi)F_{\mu\nu}\tilde{F}^{\mu\nu} +\mathcal{L}_{\rm ch}(\chi_{a},\,A_{\nu})\right],
\end{equation}
where $g={\rm det\,}g_{\mu\nu}$ is the determinant of the spacetime metric, $I(\phi)$ is the axial-coupling function, $F_{\mu\nu}=\partial_{\mu}A_{\nu}-\partial_{\nu}A_{\mu}$ is the gauge-field strength tensor, and	
\begin{equation}
	\label{F}
	\tilde{F}^{\mu\nu}=\frac{1}{2\sqrt{-g}}\,\varepsilon^{\mu\nu\lambda\rho}F_{\lambda\rho}
\end{equation}
is the corresponding dual tensor; $\varepsilon^{\mu\nu\lambda\rho}$ is the absolutely antisymmetric Levi-Civita symbol with $\varepsilon^{0123}=+1$. The last term in Eq.~(\ref{S}) corresponds to all matter fields $\chi_{a}$ charged under the $U(1)$ gauge group and, therefore, coupled to $A_{\mu}$. 

For the sake of generality, we do not specify the inflationary model and the axial-coupling function $I(\phi)$; we assume that the inflaton dynamics is known and consider the generation of gauge fields on this background. (This can be done only in the absence of the backreaction of generated fields which will be discussed below.) The equation of motion for the gauge field following from action~(\ref{S}) reads as
\begin{equation}
	\label{L_E_2}
	\frac{1}{\sqrt{-g}}\partial_{\mu}\left[\sqrt{-g}\,F^{\mu\nu} \right]+ \frac{dI}{d\phi}\,\tilde{F}^{\mu\nu}\partial_{\mu}\phi=j^{\nu},
\end{equation}
where 
\begin{equation}
	j^{\nu}=-\frac{\partial \mathcal{L}_{\rm ch}(\chi_{a},\,A_{\mu})}{\partial A_{\nu}}
\end{equation}
is the electric 4-current induced by the Schwinger effect. In addition, the Bianchi identity for the dual gauge-field strength tensor must be satisfied:
\begin{equation}
	\label{Bianchi}
	\frac{1}{\sqrt{-g}}\partial_{\mu}\left[\sqrt{-g}\,\tilde{F}^{\mu\nu} \right] = 0.
\end{equation}

To switch to a 3-vector notation, we introduce the electric $\bm{E}=(E^{1},\,E^{2},\,E^{3})$ and magnetic $\bm{B}=(B^{1},\,B^{2},\,B^{3})$ fields as follows:
\begin{equation}
F_{0i}=aE^{i}, \quad F_{ij}=-a^{2}\varepsilon_{ijk} B^{k}, \quad \tilde{F}_{0i}=aB^{i}, \quad \tilde{F}_{ij}=a^{2}\varepsilon_{ijk} E^{k}.
\end{equation}
Moreover, the electric current 4-vector can be represented as
\begin{equation}
	j^{\mu}=\big(0,\,\frac{1}{a}\bm{J}\big),
\end{equation}
where we assumed a vanishing charge density because of the quasineutrality of the plasma produced due to the Schwinger effect. Note that all 3-vectors represent physical quantities measured by a comoving observer. Then, Maxwell's equations read
\begin{equation}
	\label{Maxwell_1}
	\dot{\bm{E}}+2 H \bm{E}- [\boldsymbol{\nabla}_{\rm \!\!ph}\times\bm{B}] + 2H\xi \,\bm{B}+\bm{J}=0,
\end{equation}
\begin{equation}
	\label{Maxwell_2}
	\dot{\bm{B}}+2 H \bm{B}+[\boldsymbol{\nabla}_{\rm \!\!ph}\times\bm{E}]=0,
\end{equation}
\begin{equation}
	\label{Maxwell_3}
	\boldsymbol{\nabla}_{\rm \!\!ph}\cdot \bm{E}=0, \qquad \boldsymbol{\nabla}_{\rm \!\!ph}\cdot \bm{B}=0,
\end{equation}
where the dot over a symbol denotes its derivative with respect to time $t$,
$H\equiv H(t)=\frac{\dot a(t)}{a(t)}$ is the Hubble parameter, and $\boldsymbol{\nabla}_{\rm \!\!ph}=\partial/\partial \bm{x}_{\rm ph}=(1/a)\partial/\partial \bm{x}$ is the spatial gradient operator in physical coordinates $\bm{x}_{\rm ph}=a \bm{x}$. 
We also introduced the dimensionless parameter
\begin{equation}
	\xi=\frac{dI}{d\phi}\frac{\dot{\phi}}{2H},
\end{equation}
which controls the efficiency of gauge-field production due to the axion--vector coupling.

To close the system of Maxwell's equations, we will assume that the induced current of charged particles produced by the Schwinger effect satisfies Ohm's law,
\begin{equation}
\label{Ohms-law}
	\bm{J}=\sigma\bm{E},
\end{equation}
where $\sigma$ is the generalized conductivity, which depends only on the absolute values of electric and magnetic fields. In the case of one Dirac fermion of mass $m$ and hypercharge $Q$, the Schwinger conductivity reads
\begin{equation}
\label{sigma-fermion}
	\sigma=\frac{|g'Q|^{3}}{6\pi^{2}}\frac{|B|}{H}{\rm coth}\Big(\frac{\pi |B|}{|E|}\Big)\exp\Big(-\frac{\pi m^{2}}{|g' Q E|}\Big),
\end{equation}
where $g'$ is the SM $U(1)_{Y}$ gauge coupling constant, $|E|=\sqrt{\langle \bm{E}^{2}\rangle}$, $|B|=\sqrt{\langle \bm{B}^{2}\rangle}$, and $\langle\ldots\rangle$ denotes the vacuum expectation value.

This expression was derived in the case of constant and collinear electric and magnetic fields in de Sitter spacetime (see, e.g., Ref.~\cite{Domcke:2019}). We will utilize this approximation in our analysis, too, assuming that the electric and magnetic fields change adiabatically slowly. To be more precise, we employ Eq.~\eqref{sigma-fermion} in order to estimate the hyperelectric conductivity of the SM plasma in the presence of a hyperelectromagnetic background field,
\begin{equation}
\label{sigma-SM}
\frac{\sigma_{\mathrm{SM}}}{2H}=
a_{\rm SM}\sqrt{\frac{\langle \bm{B}^{2}\rangle}{H^{4}}}{\rm coth}\Big(\pi\sqrt{\frac{\langle \bm{B}^{2}\rangle}{\langle \bm{E}^{2}\rangle}}\Big) \,,\qquad a_{\rm SM} = \frac{41g^{\prime 3}}{144\pi^{2}} \,, 
\end{equation}
where the factor of $2H$ is moved to the left-hand side in order to obtain a dimensionless quantity, which will become convenient later on (see below). The expression in Eq.~\eqref{sigma-SM} only accounts for the production of massless SM fermions during axion inflation. In principle, the SM Higgs boson, which also interacts with the hypercharge gauge field, could be produced during axion inflation as well. However, to ensure unbroken electroweak symmetry and hence massless SM fermions, we assume that the SM Higgs field remains stabilized at the origin in field space by a large mass term throughout inflation. Such a large mass can, e.g., be induced by a nonminimal coupling to the Ricci curvature scalar $R$. In the numerical evaluation of Eq.~\eqref{sigma-SM}, specifically in the evaluation of the numerical coefficient $a_{\rm SM}$, we also take into account the energy dependence of the hypercharge gauge coupling constant~\cite{Srednicki-book},
\begin{equation}
\label{running}
\frac{1}{g^{\prime2}(\mu)} =
\frac{1}{g^{\prime2}(m_{Z})} + \frac{41}{48\pi^{2}}\ln\frac{m_{Z}}{\mu}.
\end{equation}
Here, we use the full SM beta function of $g'$; threshold effects because of the large Higgs mass during inflation are model dependent and numerically negligible. At the energy scale of the $Z$-boson mass, $m_{Z}\approx 91.2\,$GeV, the gauge coupling equals $g^{\prime}(m_{Z})\approx 0.35$. For a characteristic energy scale $\mu$ relevant for Schwinger pair production, we use
\begin{equation}
\label{eq:muscale}
\mu=\Big(\frac{\langle\bm{E}^{2}\rangle+\langle\bm{B}^{2}\rangle}{2}\Big)^{1/4}\,.
\end{equation}

In any specific model of inflation, the Hubble rate $H$ and $\xi$ are functions of time. However, if they change adiabatically slowly (which is consistent with the slow-roll regime during inflation), some order-of-magnitude estimates for the generated gauge fields can be obtained by considering the case of $H={\rm const}$ and $\xi={\rm const}$. These estimates can be then used in any other model where the same values of $H$ and $\xi$ are realized. The main purpose of the present work is to derive such model-independent estimates for a wide range of constant $H$ and $\xi$.

Handling vector quantities in position space makes the numerical analysis very demanding. That is why we will utilize the gradient-expansion formalism developed in Ref.~\cite{Gorbar:2021} for the description of hypermagnetogenesis during axion inflation. It employs the vacuum expectation values of scalar products of the electric and/or magnetic field vectors with an arbitrary number of spatial curls acting on them. In this work, it will be more convenient for us to slightly adapt our notation and introduce the following set of dimensionless  bilinear electromagnetic functions:
\begin{equation}
	\label{E_n}
	\mathcal{E}^{(n)}=\frac{1}{H^{n+4}}\left\langle \bm{E}\cdot (\boldsymbol{\nabla}_{\rm \!\!ph}\times)^{n} \bm{E}  \right\rangle,
\end{equation}
\begin{equation}
	\label{G_n}
	\mathcal{G}^{(n)}=-\frac{1}{H^{n+4}}\left\langle \bm{E}\cdot (\boldsymbol{\nabla}_{\rm \!\!ph}\times)^{n} \bm{B}  \right\rangle,
\end{equation}	
\begin{equation}
	\label{B_n}
	\mathcal{B}^{(n)}=\frac{1}{H^{n+4}}\left\langle \bm{B}\cdot (\boldsymbol{\nabla}_{\rm \!\!ph}\times)^{n} \bm{B}  \right\rangle.
\end{equation}
Now, using Maxwell's equations (\ref{Maxwell_1}) and (\ref{Maxwell_2}), we obtain the system of equations for these functions,
\begin{equation}
	\label{dot_E_n}
	\mathcal{E}^{(n)\prime} + (n+4+4s)\,	\mathcal{E}^{(n)} - 4\xi\,\mathcal{G}^{(n)} +2\mathcal{G}^{(n+1)}=[\mathcal{E}^{(n)\prime}]_{b},
\end{equation}
\begin{equation}
	\label{dot_G_n}
	\mathcal{G}^{(n)\prime} +(n+4+2s)\, \mathcal{G}^{(n)}-\mathcal{E}^{(n+1)}+\mathcal{B}^{(n+1)} - 2\xi\,\mathcal{B}^{(n)}=[\mathcal{G}^{(n)\prime}]_{b},
\end{equation}
\begin{equation}
	\label{dot_B_n}
	\mathcal{B}^{(n)\prime} + (n+4)\,	\mathcal{B}^{(n)}-2\mathcal{G}^{(n+1)}=[\mathcal{B}^{(n)\prime}]_{b},
\end{equation} 
where the prime denotes the derivative with respect to dimensionless time $\tau=Ht$ and $s\equiv s(\tau)=\frac{\sigma(\tau)}{2H}$ is the dimensionless conductivity. Terms on the right-hand sides of these equations are the boundary terms, which take into account that the number of physically relevant gauge-field modes outside the horizon continuously grows in time during inflation. Indeed, in Ref.~\cite{Gorbar:2021}, it is shown that the momentum $k_{\rm h}$ of the mode crossing the horizon [defined in such a way that all modes with $k<k_{\rm h}(t)$ have already experienced the tachyonic instability at time $t$] changes in time as
\begin{equation}
	k_{\rm h}=aH(|\xi|+\sqrt{\xi^{2}+s^{2}+s}).
\end{equation}
Since $a(t)$ exponentially grows during inflation, new modes cross the horizon, undergo the quantum-to-classical transition and start contributing to the classical gauge field.

The explicit expressions for the boundary terms were derived in Ref.~\cite{Gorbar:2021}. For the dimensionless quantities introduced in Eqs.~(\ref{E_n})--(\ref{B_n}), they take the form
\begin{equation}
	\label{E_p_d}
	[\mathcal{E}^{(n)\prime}]_{b}=\frac{\Delta}{4\pi^{2}}[r(\xi,\,s)]^{n+4}\sum_{\lambda=\pm 1}\lambda^{n} E_{\lambda}(\xi,\,s),
\end{equation}
\begin{equation}
	\label{G_p_d}
	[\mathcal{G}^{(n)\prime}]_{b}=\frac{\Delta}{4\pi^{2}}[r(\xi,\,s)]^{n+4}\sum_{\lambda=\pm 1}\lambda^{n+1}G_{\lambda}(\xi,\,s),
\end{equation}
\begin{equation}
	\label{B_p_d}
	[\mathcal{B}^{(n)\prime}]_{b}=\frac{\Delta}{4\pi^{2}}[r(\xi,\,s)]^{n+4}\sum_{\lambda=\pm 1}\lambda^{n}B_{\lambda}(\xi,\,s),
\end{equation}
where
\begin{equation}
	\label{E-lambda}
	E_{\lambda}(\xi,s)=\frac{e^{\pi\lambda \xi}}{r^{2}(\xi,s)} \left|\left(i r(\xi,s)-i\lambda \xi-s\right)W_{-i\lambda\xi,\frac{1}{2}+s}(-2i r(\xi,s))+W_{1-i\lambda\xi,\frac{1}{2}+s}(-2i r(\xi,s))\right|^{2},
\end{equation}
\begin{equation}
	\label{G-lambda}
	G_{\lambda}(\xi,s)=
	\frac{e^{\pi\lambda \xi}}{r(\xi,s)} \left\{\Re e\left[W_{i\lambda \xi,\frac{1}{2}+s}(2i r(\xi,s)) W_{1-i\lambda\xi,\frac{1}{2}+s}(-2i r(\xi,s))\right]-s\left|W_{-i\lambda\xi,\frac{1}{2}+s}(-2i r(\xi,s)) \right|^{2}\right\},
\end{equation}
\begin{equation}
	\label{B-lambda}
	\qquad B_{\lambda}(\xi,s)=e^{\pi\lambda \xi}\,\left|W_{-i\lambda\xi,\frac{1}{2}+s}(-2i r(\xi,s)) \right|^{2}.
\end{equation}
Here, $W_{-i\lambda\xi,\frac{1}{2}+s}$ is the Whittaker function, and $r(\xi,s)=|\xi|+\sqrt{\xi^{2}+s+s^{2}}$ is the dimensionless physical momentum of the horizon-crossing mode.

We would also like to highlight the parameter $\Delta$ in Eqs.~(\ref{E_p_d})--(\ref{B_p_d}), which was recently discussed for the first time in Ref.~\cite{Gorbar:2021} and which modulates the magnitude of the boundary terms. It originates from the fact that gauge-field vacuum fluctuations inside the horizon are damped in the conducting medium. Indeed, as is shown in Ref.~\cite{Gorbar:2021}, the mode function deep inside the horizon (i.e., for $k\gg k_{\rm h}$) is represented by the damped Bunch-Davies vacuum
\begin{equation}
\label{eq:BD}
A_{\lambda}(t,\bm{k})=\sqrt{\frac{\Delta(t)}{2k}}e^{-i k \eta(t)},
\end{equation}
where $\eta=\int^{t} dt'/a(t')$ is the conformal time and
\begin{equation}
\label{eq:delta}
\Delta(t)=\exp\Big(-\int\limits_{-\infty}^{t}\sigma(t')dt'\Big)
\end{equation}
is the damping factor which depends on the conductivity at times $t' \leq t$ and thus makes the gauge-field evolution inherently nonlocal in time. Note that, in Eq.~\eqref{eq:delta}, we integrate over $t'$ up to the infinite past $t'\rightarrow -\infty$. This implies that nonzero conductivity at some $t'$ results in a suppression of \textit{all} subhorizon gauge-field modes up to arbitrarily large $k$ values, even modes that are deep inside the Hubble horizon
at time $t'$ and which only become tachyonically unstable at times much later than $t'$. To some extent, this is an approximation and technical simplification, as we expect the range of $k$ values affected by the electric conductivity on subhorizon scales to be finite. Gauge-field modes with momenta much larger than the momenta of the charged fermions in the plasma should, e.g., not suffer from the damping induced by the nonvanishing conductivity. In principle, the lower integration boundary in Eq.~\eqref{eq:delta} should therefore be replaced by a finite $k$-dependent cutoff $t_{\rm UV}\left(k\right)$ ensuring that $\Delta$ only receives contributions from times $t' \geq t_{\rm UV}\left(k\right)$ when the gauge-field mode with momentum $k$ has a spatial extent larger than some UV length scale. However, at present, no exact expression for $t_{\rm UV}\left(k\right)$, is known. In the following, we will therefore stick to the standard approach in the literature and simply treat all $k$ modes on an equal footing. In particular, we do not attempt to determine the UV cutoff scale that is expected to enter Eqs.~\eqref{eq:BD} and \eqref{eq:delta} at some point. This task requires further investigation and is left for future work. For the purposes of this work, it suffices to note that the conductivity $\sigma$ is often a monotonically (even exponentially) increasing function of time that reaches its largest value toward the end of inflation. In realistic scenarios, the integral in Eq.~\eqref{eq:delta} is therefore typically dominated by the upper integration boundary, which drastically reduces the sensitivity to the lower integration boundary.

Even without a more sophisticated treatment of the lower integration boundary in Eq.~\eqref{eq:delta}, the dependence of the boundary terms on $\Delta$ complicates our model-independent analysis for several reasons. For instance, if two models exhibit the same values of $H$ and $\xi$, they could still have different values of $\Delta$ because of a different prehistory. Moreover, even though the parameters $H$ and $\xi$ can be constant (or changing adiabatically slowly) in a real situation, the parameter $\Delta$ always decreases in time unless the conductivity vanishes. Finally, in contrast to $H$ and $\xi$, \textit{a priori} it is difficult to estimate the value of the parameter $\Delta$ without solving the full self-consistent system of equations for the inflaton, gauge fields, and charged particles. Nevertheless, we will show in the subsequent section that the magnitude of the generated field, in the case when the Schwinger effect is important, depends only weakly on $\Delta$. Therefore, one does not necessarily need to know the exact value of $\Delta$; a rough estimate suffices. This serves as an other reason why we defer a more detailed investigation of the lower integration boundary in Eq.~\eqref{eq:delta} to future work. 

The system of equations (\ref{dot_E_n})--(\ref{dot_B_n}) for the gradient-expansion formalism is infinite by construction, since the equation for the quantity of order $n$ contains quantities of order $(n+1)$. However, there exists a simple approximation, allowing us to truncate this chain at some finite order. Indeed, for large enough $n$, the dominant contributions to the quantities $\mathcal{E}^{(n)}$, $\mathcal{B}^{(n)}$, and $\mathcal{G}^{(n)}$ correspond to the shortest modes in their spectra, i.e., modes in the vicinity of the horizon crossing mode $k_{\rm h}$. This allows us to express the higher-order quantities in terms of the lower-order ones \cite{Gorbar:2021}, e.g.,
\begin{equation}
	\mathcal{E}^{(n+1)}\approx r(\xi,\, s)^{2} \mathcal{E}^{(n-1)}
\end{equation}
and so on. Applying these relations for some $n=n_{\rm max}$, one can truncate the infinite system of equations (\ref{dot_E_n})--(\ref{dot_B_n}).

\section{Numerical analysis}
\label{sec-numerical}

\begin{figure}[ht!]
\centering
\includegraphics[width=0.97\textwidth]{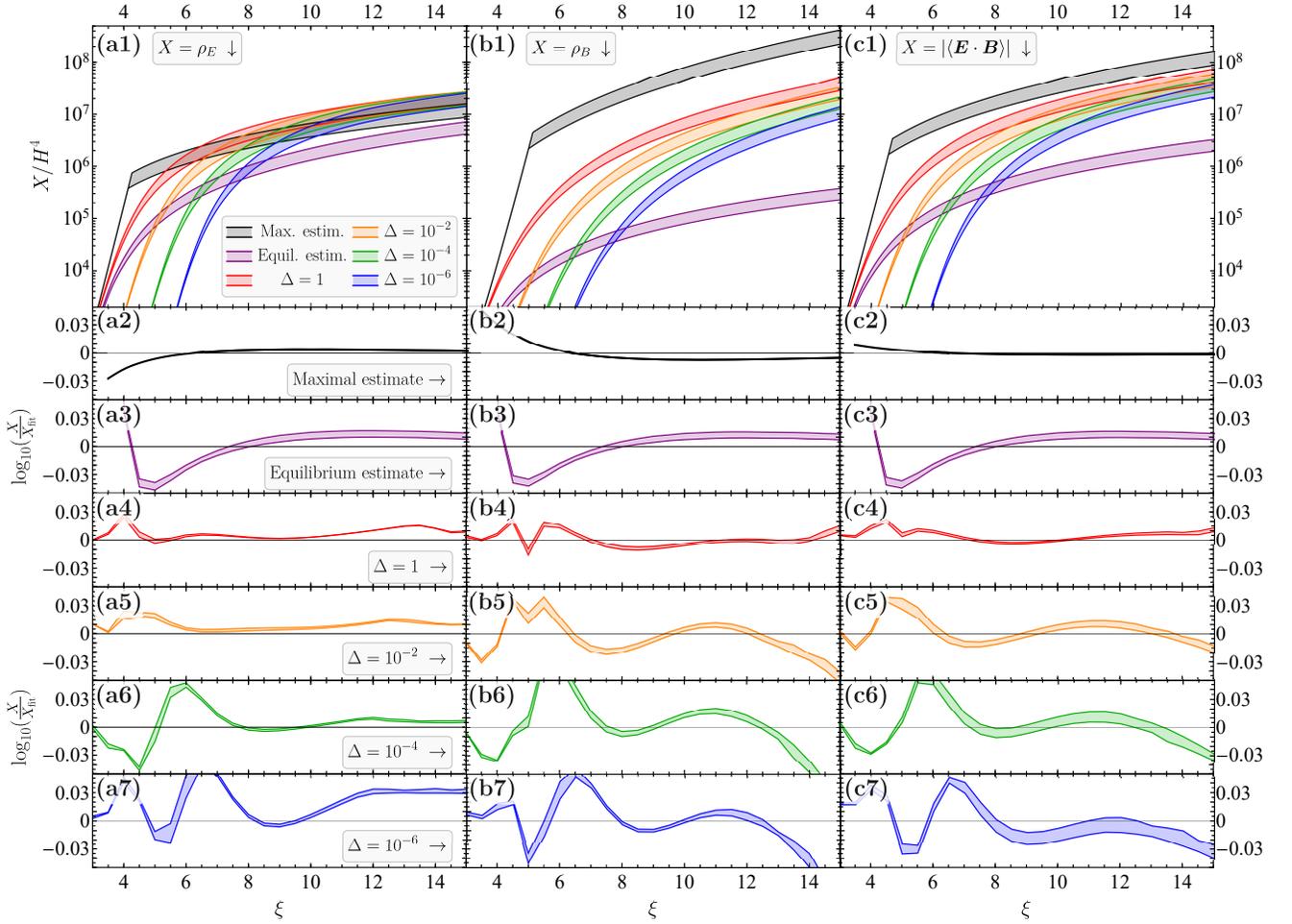}
\caption{Model-independent results for the efficiency of hypermagnetogenesis during axion inflation, specifically, for the hyperelectric energy density $\rho_E$ (column a), hypermagnetic energy density $\rho_B$ (column b), and hypercharge Chern-Pontryagin density $\left|\langle\bm{E}\cdot\bm{B}\rangle\right|$ (column c), in dependence on the gauge-field production parameter $\xi$, Hubble rate $H$, and damping factor $\Delta$. The Hubble rate is varied between $10^8$ and $10^{14}\,\textrm{GeV}$; the dependence on $H$ is shown in the form of bands whose upper (lower) edges correspond to $H = 10^8\,\textrm{GeV}$ ($H = 10^{14}\,\textrm{GeV}$). The gray and purple bands in the first row, respectively, correspond to the maximal and equilibrium estimates derived in Refs.~\cite{Domcke:2018,Domcke:2019bar} (see Secs.~\ref{subsec:maxestimate} and \ref{subsec:eqestimate}). All other colorful bands in the first row correspond to our new results based on the gradient-expansion formalism. For all three types of estimates, we derive fit functions; in rows 2 to 7, we compare these fit functions to the exact numerical results in row 1.}
\label{fig-compare}
\end{figure}

The gradient-expansion formalism outlined in the previous section now allows us to determine model-independent estimates for the efficiency of gauge-field production during axion inflation. To this end, we need to fix the values of the parameters $H$, $\xi$, and $\Delta$ and look for the stationary solution of the system of equations (\ref{dot_E_n})--(\ref{dot_B_n}). We repeat this analysis for a large set of parameter points in the three-dimensional parameter space spanned by $H$, $\xi$, and $\Delta$ and present our numerical results for the electric and magnetic energy densities $\rho_E$ and $\rho_B$ as well as for $\left|\langle \bm{E}\cdot\bm{B}\rangle\right|$ in Fig.~\ref{fig-compare}. In the following, we will compare our numerical results to existing estimates of $\rho_E$, $\rho_B$, and $\left|\langle\bm{E}\cdot\bm{B}\rangle\right|$ in the literature~\cite{Domcke:2018,Domcke:2019bar} (see Secs.~\ref{subsec:maxestimate} and \ref{subsec:eqestimate}) and construct semianalytical fit functions for these old estimates as well as for our own new results, i.e., for all functions shown in the first row of Fig.~\ref{fig-compare} (see Sec.~\ref{subsec:fits}). Rows 2 to 7 in Fig.~\ref{fig-compare} show the differences between these fit functions and the corresponding exact numerical results in row 1, which clearly demonstrates the high accuracy of our fit functions. Based on these fit functions, it is therefore now possible to reconstruct and utilize all existing estimates of gauge-field production during axion inflation, including the estimates in Refs.~\cite{Domcke:2018,Domcke:2019bar} as well as our new estimates based on the gradient-expansion formalism, without any further numerical analysis. Our fit functions provide all the necessary information in a compact and ready-to-use form for future applications. Finally, to validate our results, we will compare all estimates of $\rho_E$, $\rho_B$, and $\left|\langle\bm{E}\cdot\bm{B}\rangle\right|$ to the exact numerical results in a specific inflationary model, namely, $m^2\phi^2/2$ inflation, in Sec.~\ref{subsec:validation}. This will lead us to the conclusion that our new model-independent results can reproduce the exact results in a given model within an order of magnitude or so as well as that our new estimates typically improve over the existing estimates. 

\subsection{Maximal estimate}
\label{subsec:maxestimate}

The authors of Refs.~\cite{Domcke:2018,Domcke:2019bar} account for the effect of fermion production during axion inflation in terms of an effective gauge-field production parameter $\xi_{\rm eff}$. This is possible in the limit of perfectly parallel or antiparallel electric and magnetic fields, in which the induced current $\bm{J}$ can also be expressed in terms of the magnetic field,
\begin{equation}
\bm{J} = \sigma\bm{E} = \textrm{sgn}\left(\bm{E}\cdot\bm{B}\right)\,\frac{\sigma\left|E\right|}{\left|B\right|}\,\bm{B} \,,\qquad \textrm{sgn}\left(\bm{E}\cdot\bm{B}\right) = - \textrm{sgn}\left(\xi\right) \,,
\end{equation}
where the relation between $\textrm{sgn}\left(\bm{E}\cdot\bm{B}\right)$ and $\textrm{sgn}\left(\xi\right)$ follows from our sign convention in the axion--vector coupling in Eq.~\eqref{S}. Making use of this relation, Amp\`ere's law in the axion-inflationary background [see Eq.~\eqref{Maxwell_1}] reads
\begin{equation}
\label{eq:ampere}
\dot{\bm{E}}+2 H \bm{E}- [\boldsymbol{\nabla}_{\rm \!\!ph}\times\bm{B}] + 2H\xi_{\rm eff} \,\bm{B} =0 \,,\qquad 
\end{equation}
with the effective gauge-field production parameter $\xi_{\rm eff}$ being defined as 
\begin{equation}
\label{eq:xieff}
\left|\xi_{\rm eff}\right| = \left|\xi\right| - \frac{\sigma\left|E\right|}{2\left|B\right|H} = \left|\xi\right| - a_{\rm SM} \coth\left(\frac{\pi\left|B\right|}{\left|E\right|}\right)\frac{\left|E\right|}{H^2} \,.
\end{equation}
The same effective parameter also appears in the equation that describes the time evolution of the energy density stored in the hyperelectromagnetic field~\cite{Gorbar:2021},
\begin{equation}
\dot{\rho}_{\rm em} + 4 H\rho_{\rm em} =  \left[\dot{\rho}_{\rm em}\right]_b - 2\xi\left<\bm{E}\cdot\bm{B}\right>H - \sigma \langle\bm{E}^2\rangle \,, \qquad \rho_{\rm em} = \rho_E + \rho_B = \frac{1}{2}\left(\langle\bm{E}^2\rangle+\langle\bm{B}^2\rangle\right) \,,
\end{equation}
which, in the limit of parallel or antiparallel electric and magnetic fields, turns into
\begin{equation}
\label{eq:rhoEM}
\dot{\rho}_{\rm em} + 4 H\rho_{\rm em} = \left[\dot{\rho}_{\rm em}\right]_b + 2\left|\xi_{\rm eff}\right|\left|E\right|\left|B\right| H   \,.
\end{equation}
Based on Eqs.~\eqref{eq:ampere} and \eqref{eq:rhoEM}, the authors of Refs.~\cite{Domcke:2018,Domcke:2019bar} now construct two estimates for the electric and magnetic fields generated during axion inflation. We will first discuss the estimate based on Eq.~\eqref{eq:rhoEM}, which we will refer to as the ``maximal estimate'' in the following, for reasons that will become clear shortly. 

Let us consider the idealized situation of a stationary de Sitter background with constant values of $\xi$ and $H$. In such a background, we expect the electric and magnetic field strengths to reach a stationary attractor solution that is solely described by the values of $\xi$ and $H$.%
\footnote{We expect that, in a stationary de Sitter background with constant $\xi$ and $H$, a refined treatment of the lower integration boundary in Eq.~\eqref{eq:delta} will turn $\Delta$ from an independent parameter into a dependent parameter, $\Delta = \Delta \left(\xi,H\right)$; see the discussion below Eq.~\eqref{eq:delta}.}
This solution will therefore be independent of time, which allows us to set $\dot{\rho}_{\rm em} = 0$ in Eq.~\eqref{eq:rhoEM}. Moreover, if we momentarily interpret $\rho_{\rm em}$ as the energy density stored in the entire hyperelectromagnetic field on super- and subhorizon scales and not only stored in the gauge-field modes that have already become tachyonically unstable, we can also drop the boundary term in Eq.~\eqref{eq:rhoEM},
\begin{equation}
\label{eq:consistency}
4 H\rho_{\rm em} = 2\left|\xi_{\rm eff}\right|\left|E\right|\left|B\right| H   \,.
\end{equation}
This relation represents a consistency condition, based on the requirement of energy conservation, that applies to any stationary solution for the electric and magnetic fields.  In the context of our gradient-expansion formalism and introducing the shorthand notation $\mathcal{E} \equiv \mathcal{E}^{(0)}$, $\mathcal{B} \equiv \mathcal{B}^{(0)}$, $\mathcal{G} \equiv \mathcal{G}^{(0)}$, the condition in  Eq.~\eqref{eq:consistency} obtains the form
\begin{equation}
\label{eq:consistency2}
\mathcal{E} + \mathcal{B} - \left|\xi_{\rm eff}\right| \left|\mathcal{G}\right| = 0 \,,\qquad \left|\xi_{\rm eff}\right| = \left|\xi\right| - a_{\rm SM} \coth\left(\pi\sqrt{\mathcal{B}/\mathcal{E}}\right)\sqrt{\mathcal{E}} \,,
\end{equation}
which can be solved for $\left|\mathcal{G}\right| \equiv \left|\left<\bm{E}\cdot\bm{B}\right>\right|/H^4$ as a function of the ratio of the electric and magnetic field strengths, $\mathcal{R}$,
\begin{equation}
\label{eq:GR}
\left|\mathcal{G}\right|_{\mathcal{R}} = \frac{1}{a_{\rm SM}^2\mathcal{R}^3}\left(1 + \mathcal{R}^2 - \left|\xi\right|\mathcal{R}\right)^2\tanh^2\left(\pi/\mathcal{R}\right) \,, \qquad \mathcal{R} = \frac{\left|E\right|}{\left|B\right|} = \sqrt{\frac{\mathcal{E}}{\mathcal{B}}} \,.
\end{equation}
where $\mathcal{R}$ takes values in the range
\begin{equation}
\frac{1}{2}\left(\left|\xi\right|-\sqrt{\left|\xi\right|^2 -4}\right) \leq \mathcal{R} \leq  \frac{1}{2}\left(\left|\xi\right|+\sqrt{\left|\xi\right|^2 -4}\right) \,.
\end{equation}
Outside this interval, no solution of the consistency condition in Eq.~\eqref{eq:consistency2} exists. Any attractor solution with $\dot{\rho}_{\rm em} = 0$ satisfies the relation between $\left|\mathcal{G}\right|$ and $\mathcal{R}$ in Eq.~\eqref{eq:GR}, which means in particular that any attractor solution for $\left|\mathcal{G}\right|$ is always bounded from above by the maximal value that $\left|\mathcal{G}\right|_{\mathcal{R}}$ can obtain as a function of  $\mathcal{R}$,
\begin{equation}
\label{eq:Gmax}
\left|\mathcal{G}\right|_{\rm max} = \max_{\mathcal{R}} \left|\mathcal{G}\right|_{\mathcal{R}} \,.
\end{equation}
Because of the nontrivial functional dependence of $\left|\mathcal{G}\right|_{\mathcal{R}}$ on $\mathcal{R}$, it is unfortunately not possible to write down a closed analytical expression for $\left|\mathcal{G}\right|_{\rm max}$. Instead, we need to maximize $\left|\mathcal{G}\right|_{\mathcal{R}}$, for fixed $\left|\xi\right|$, over the admissible range of $\mathcal{R}$ values numerically. Moreover, once we know the value $\mathcal{R}_{\rm max}$ that maximizes $\left|\mathcal{G}\right|_{\mathcal{R}}$ for a given $\left|\xi\right|$, we can use it to determine the corresponding values of $\mathcal{E}$ and $\mathcal{B}$,
\begin{equation}
\label{eq:EBmax}
\mathcal{E}_{\rm max} = \mathcal{R}_{\rm max}\left|\mathcal{G}\right|_{\rm max} \,, \qquad \mathcal{B}_{\rm max} = \frac{\left|\mathcal{G}\right|_{\rm max} }{\mathcal{R}_{\rm max}}\,.
\end{equation}
We caution that $\mathcal{E}_{\rm max}$ and $\mathcal{B}_{\rm max}$ do not correspond to the maximal values of $\mathcal{E}$ and $\mathcal{B}$ that are consistent with the condition in Eq.~\eqref{eq:consistency2}; they rather correspond to the pair of $\left|\mathcal{G}\right|$ and $\mathcal{R}$ values that maximize $\left|\mathcal{G}\right|_{\mathcal{R}}$. Together, $\mathcal{E}_{\rm max}$, $\mathcal{B}_{\rm max}$, and $\left|\mathcal{G}\right|_{\rm max}$ represent what we will refer to as the maximal estimate in the following. Our numerical results for the maximal estimate, based on the maximization described in Eq.~\eqref{eq:Gmax}, are shown by the gray bands in the first row in Fig.~\ref{fig-compare}. The dependence on the Hubble rate in these results enters through the running of the hypercharge gauge coupling constant inside the factor $a_{\rm SM}$, which we determine in a self-consistent manner so as to satisfy the relations in Eqs.~\eqref{running} and \eqref{eq:muscale}. As can be seen in Fig.~\ref{fig-compare}, the gray bands extend down to $\left|\xi\right| \sim 4 - 5$. At lower values of $\left|\xi\right|$, the numerical results for $\mathcal{E}_{\rm max}$, $\mathcal{B}_{\rm max}$, and $\left|\mathcal{G}\right|_{\rm max}$ begin to exceed the corresponding quantities in the free case without any backreaction or fermion production, which is physically not motivated. The effect of fermion production should always suppress the efficiency of gauge-field production and not enhance it. At low $\left|\xi\right|$ values, we therefore show the free solutions for $\mathcal{E}$, $\mathcal{B}$, and $\left|\mathcal{G}\right|$ rather than the numerical output of Eqs.~\eqref{eq:Gmax} and \eqref{eq:EBmax}.

In the following, we will now discuss simple and novel fit functions that manage to describe our exact numerical results for  $\mathcal{E}_{\rm max}$, $\mathcal{B}_{\rm max}$, and $\left|\mathcal{G}\right|_{\rm max}$ to high accuracy. The starting point of our construction is going to be the value of $\left|\mathcal{G}\right|_{\mathcal{R}}$ evaluated at $\mathcal{R} = 1$, which corresponds to equal amounts of energy in the electric and magnetic fields,
\begin{equation} 
\label{eq:GR1}
\left|\mathcal{G}\right|_{\mathcal{R}=1} = \frac{1}{a_{\rm SM}^2}\left(\left|\xi\right|-2\right)^2\tanh^2\left(\pi\right) \,.
\end{equation}
It then turns out that our numerical data are well described by the fit functions
\begin{align}
\label{eq:Emax}
\mathcal{E}_{\rm max} & \simeq \left(1.5922 + 0.4561\left|\xi\right|\right)\left|\xi\right|^{-1} \left|\mathcal{G}\right|_{\mathcal{R}=1} \,, \\
\label{eq:Bmax}
\mathcal{B}_{\rm max} & \simeq \left(0.2706 + 0.0472\left|\xi\right|\right)\left|\xi\right|^{+1} \left|\mathcal{G}\right|_{\mathcal{R}=1} \,, \\
\label{eq:Gmx}
\left|\mathcal{G}\right|_{\rm max} & \simeq \left(0.6637 + 0.1472\left|\xi\right|\right) \left|\xi\right|^{0}\ \  \left|\mathcal{G}\right|_{\mathcal{R}=1} \,.
\end{align}
In addition to $\left|\xi\right|$, these three functions also depend on $H$ through the energy dependence of the hypercharge gauge coupling constant. To the first approximation, this effect may be neglected, and $g'$ may be simply set to a characteristic value around $g' \sim 0.4$. However, to obtain even more accurate estimates, we are actually able to determine the self-consistent value of $g'$ analytically. To this end, we need to express the renormalization scale $\mu$ in Eq.~\eqref{eq:muscale} in terms of the semianalytical expressions for $\mathcal{E}_{\rm max}$ and $\mathcal{B}_{\rm max}$ and factor out the dependence on the gauge coupling constant $g'$,
\begin{equation}
\mu = \frac{H}{2^{1/4}}\left(\mathcal{E}_{\rm max} + \mathcal{B}_{\rm max}\right)^{1/4} = \frac{\bar{\mu}_{\rm max}}{g'^{3/2}} \,,
\end{equation}
where the rescaled quantity $\bar{\mu}_{\rm max}$,
\begin{equation}
\bar{\mu}_{\rm max} = \frac{H}{2^{1/4}}\left(\overline{\mathcal{E}}_{\rm max} + \overline{\mathcal{B}}_{\rm max}\right)^{1/4} \,, 
\end{equation}
is defined in terms of rescaled versions of $\mathcal{E}_{\rm max}$ and $\mathcal{B}_{\rm max}$ that no longer depend on $g'$,
\begin{align}
\overline{\mathcal{E}}_{\rm max} & \simeq \frac{1}{\bar{a}_{\rm SM}^2} \left(1.5922 + 0.4561\left|\xi\right|\right)\left|\xi\right|^{-1} \left(\left|\xi\right|-2\right)^2\tanh^2\left(\pi\right) \,, \\
\overline{\mathcal{B}}_{\rm max} & \simeq \frac{1}{\bar{a}_{\rm SM}^2} \left(0.2706 + 0.0472\left|\xi\right|\right)\left|\xi\right|^{+1} \left(\left|\xi\right|-2\right)^2\tanh^2\left(\pi\right) \,,
\end{align}
because they are no longer proportional to $a_{\rm SM}$ but instead proportional to 
\begin{equation}
\bar{a}_{\rm SM} = \frac{a_{\rm SM}}{g'^3} = \frac{41}{144\pi^2} \,.
\end{equation}
Making use of these definitions, the hypercharge gauge coupling constant in Eq.~\eqref{running} can be written as
\begin{equation}
\frac{1}{g'^2} = \frac{1}{g'^2\left(m_Z\right)} + \frac{41}{48\pi^2}\ln\left(g'^{3/2}\,\frac{m_Z}{\bar{\mu}_{\rm max}}\right) \,.
\end{equation}
The self-consistent solution for $g'$ at the one-loop level thus follows from solving this relation for $g'$,
\begin{equation}
\label{eq:gmax}
g_{\rm max}' = \frac{1}{\sqrt{b\,W_0\left(z_{\rm max}\right)}} \,,
\end{equation}
where the subscript ``max'' again indicates that this expression for $g'$ corresponds to the self-consistent solution in the case of the maximal estimate. $W_0$ denotes the principal branch of the Lambert $W$ function or product logarithm, and the argument of the Lambert $W$ function, $z_{\rm max}$, and the numerical coefficient $b$ are given as follows:
\begin{equation}
z_{\rm max} = \frac{1}{b}\,\exp\left(\frac{1}{b\,g'^2\left(m_Z\right)}\right)\left(\frac{m_Z}{\bar{\mu}_{\rm max}}\right)^{4/3} \,, \qquad b = \frac{41}{64\pi^2} \,.
\end{equation}
Numerically, $z_{\rm max}$ evaluates to
\begin{equation}
z_{\rm max} \simeq \frac{3.6232 \times 10^{41}}{\left(6.3212/|\xi| - 4.5105 + 0.8441\,|\xi| - 0.4344\,|\xi|^2 + 0.0812\,|\xi|^3 + 0.0468\,|\xi|^4\right)^{1/3}} \left(\frac{10^{12}\,\textrm{GeV}}{H}\right)^{4/3} \,,
\end{equation}
which makes the dependence on the value of the Hubble rate $H$ explicit. If we now use the result in Eq.~\eqref{eq:gmax} in order to evaluate $a_{\rm SM}$ in Eqs.~\eqref{eq:GR1}, \eqref{eq:Emax}, \eqref{eq:Bmax}, and \eqref{eq:Gmx}, we obtain semianalytical fit functions for $\mathcal{E}_{\rm max}$, $\mathcal{B}_{\rm max}$, and $\left|\mathcal{G}\right|_{\rm max}$ that self-consistently account for the running of the gauge coupling constant. In row 2 of Fig.~\ref{fig-compare}, we compare these fit functions to the exact numerical result indicated by the three gray bands in row 1 and find excellent agreement.

Finally, before we move on to the next estimate, which we will refer to as the ``equilibrium estimate'', we mention that $g'_{\rm max}$ in Eq.~\eqref{eq:gmax} is also well approximated by the following, much simpler expression:
\begin{equation}
\label{eq:gmaxfit}
g_{\rm max}' \simeq 0.4162 + 0.0068 \log_{10} H_{12} + 0.0030 \ln\left(\frac{\left|\xi\right|}{10}\right) \,, \qquad H_{12} \equiv \frac{H}{10^{12}\,\textrm{GeV}} \,.
\end{equation}
Using this expression in Eqs.~\eqref{eq:GR1}, \eqref{eq:Emax}, \eqref{eq:Bmax}, and \eqref{eq:Gmx} results in nearly equally accurate fit functions.

\subsection{Equilibrium estimate}
\label{subsec:eqestimate}

Next, we turn to the second estimate proposed by the authors of Refs.~\cite{Domcke:2018,Domcke:2019bar}, the equilibrium estimate, which is based on the modified version of Amp\`ere's law in Eq.~\eqref{eq:ampere}. The maximal estimate constructed in the previous section represents an upper bound on all attractor solutions in a stationary de Sitter background with constant $\xi$ and $H$. The equilibrium estimate, by contrast, aims at actually constructing an explicit attractor solution. The basic idea is that, once the electric and magnetic fields have reached the attractor solution, Amp\`ere's law in Eq.~\eqref{eq:ampere} will have the same solution as in the free case without any backreaction or fermion production, the only difference being that the parameter $\xi$ in this solution needs to be replaced by $\xi_{\rm eff}$. If this is the case, the equations of motion for the gauge-field modes in Fourier space will also be solved by the usual Whittaker functions, with $\xi \rightarrow \xi_{\rm eff}$, such that  
\begin{equation}
\label{eq:Eintegral}
\mathcal{E}^{(n)} = \sum\limits_{\lambda=\pm}\frac{\lambda^{n}}{4\pi^{2}}\int\limits_{0}^{2|\xi_{\rm eff}|}dx\,x^{n+1}e^{\pi\lambda\xi_{\rm eff}}\Big|i(x-\lambda\xi_{\rm eff})W_{-i\lambda\xi_{\rm eff},\frac{1}{2}}(-2ix)+W_{1-i\lambda\xi_{\rm eff},\frac{1}{2}}(-2ix)\Big|^{2} \,,
\end{equation}
\begin{equation}
\label{eq:Gintegral}
\mathcal{G}^{(n)} = \sum\limits_{\lambda=\pm}\frac{\lambda^{n+1}}{4\pi^{2}}\!\!\!\!\int\limits_{0}^{2|\xi_{\rm eff}|}dx\,x^{n+2}e^{\pi\lambda\xi_{\rm eff}} \Re e\Big[W_{i\lambda\xi_{\rm eff},\frac{1}{2}}(2ix)W_{1-i\lambda\xi_{\rm eff},\frac{1}{2}}(-2ix)\Big] \,,
\end{equation}
\begin{equation}
\label{eq:Bintegral}
\mathcal{B}^{(n)} = \sum\limits_{\lambda=\pm}\frac{\lambda^{n}}{4\pi^{2}}\int\limits_{0}^{2|\xi_{\rm eff}|}dx\,x^{n+3}e^{\pi\lambda\xi_{\rm eff}}\Big|W_{-i\lambda\xi_{\rm eff},\frac{1}{2}}(-2ix)\Big|^{2} \,.
\end{equation}
Thus, an underlying assumption of this approach is that, after a sufficiently long time, the system reaches an attractor solution that resembles the free solution (modulo the replacement $\xi \rightarrow \xi_{\rm eff}$) not only at the level of the integrated quantities $\mathcal{E}$, $\mathcal{B}$, and $\left|\mathcal{G}\right|$ but also at the level of the Fourier mode spectrum. For large values of the effective gauge-field production parameter, $\left|\xi_{\rm eff}\right| \gg 3$, the quantities in Eqs.~\eqref{eq:Eintegral}, \eqref{eq:Bintegral}, and \eqref{eq:Gintegral}, for $n=0$, can in particular be written as
\begin{equation}
\label{eq:EBGeq}
\mathcal{E}_{\rm eq} = C_\mathcal{E}\,\frac{e^{2\pi\left|\xi_{\rm eff}\right|}}{\left|\xi_{\rm eff}\right|^3} \,, \qquad \mathcal{B}_{\rm eq} = C_\mathcal{B}\,\frac{e^{2\pi\left|\xi_{\rm eff}\right|}}{\left|\xi_{\rm eff}\right|^5} \,, \qquad \left|\mathcal{G}\right|_{\rm eq} = C_\mathcal{G}\,\frac{e^{2\pi\left|\xi_{\rm eff}\right|}}{\left|\xi_{\rm eff}\right|^4} \,,
\end{equation}
where the numerical coefficients are roughly given by $C_{\mathcal{E}} \simeq 2.6 \times 10^{-4}$, $C_{\mathcal{B}} \simeq 3.0 \times 10^{-4}$, and $C_{\mathcal{G}} \simeq \sqrt{C_{\mathcal{E}}C_{\mathcal{B}}} \simeq 2.8 \times 10^{-4}$. In the following, we will explicitly work with the relation $C_{\mathcal{G}} \simeq \sqrt{C_{\mathcal{E}}C_{\mathcal{B}}}$, which is valid in the limit of parallel or antiparallel electric and magnetic fields, i.e., the limit that has been used in the derivation of the induced current. 

Equation~\eqref{eq:EBGeq} implicitly defines the equilibrium estimate for $\mathcal{E}$, $\mathcal{B}$, and $\left|\mathcal{G}\right|$. The evaluation of this estimate is, however, complicated by the fact that the effective parameter $\xi_{\rm eff}$ in Eq.~\eqref{eq:EBGeq} also depends on the electric and magnetic field strengths. Again, we are thus not able to write down a closed analytical solution but have to resort to a numerical approach. To this end, we first note that Eq.~\eqref{eq:EBGeq} results in a simple relation between $\mathcal{R}$ and $\left|\xi_{\rm eff}\right|$,
\begin{equation}
\mathcal{R}_{\rm eq} = \sqrt{\frac{C_{\mathcal{E}}}{C_{\mathcal{B}}}}\,\left|\xi_{\rm eff}\right| \,.
\end{equation}
Using the definition of the effective gauge-field production parameter $\left|\xi_{\rm eff}\right|$ in Eq.~\eqref{eq:xieff}, we are therefore able to write
\begin{equation}
\mathcal{R}_{\rm eq} = \sqrt{\frac{C_{\mathcal{E}}}{C_{\mathcal{B}}}} \left[\left|\xi\right| - a_{\rm SM} \coth\left(\frac{\pi}{\mathcal{R}_{\rm eq}}\right)\mathcal{E}_{\rm eq}^{1/2}\right] \,,
\end{equation}
which can be solved for $\mathcal{E}_{\rm eq}$ as a function of $\left|\xi_{\rm eff}\right| = \sqrt{C_{\mathcal{B}}/C_{\mathcal{E}}}\, \mathcal{R}_{\rm eq}$. Together with the relations [see Eq.~\eqref{eq:EBGeq}]
\begin{equation}
\mathcal{B}_{\rm eq} = \frac{C_\mathcal{B}}{C_\mathcal{E}} \frac{\mathcal{E}_{\rm eq}}{\left|\xi_{\rm eff}\right|^2} \,, \qquad \left|\mathcal{G}\right|_{\rm eq} = \sqrt{\frac{C_{\mathcal{B}}}{C_{\mathcal{E}}}}\, \frac{\mathcal{E}_{\rm eq}}{\left|\xi_{\rm eff}\right|} \,,
\end{equation}
we thus find
\begin{align}
\mathcal{E}_{\rm eq} = \frac{1}{a_{\rm SM}^2}\left(\left|\xi\right| - \left|\xi_{\rm eff}\right|\right)^2\tanh^2\left(\sqrt{\frac{C_{\mathcal{B}}}{C_{\mathcal{E}}}}\,\frac{\pi}{\left|\xi_{\rm eff}\right|}\right) &  \,, \\
\mathcal{B}_{\rm eq} = \frac{C_{\mathcal{B}}/C_{\mathcal{E}}}{a_{\rm SM}^2\left|\xi_{\rm eff}\right|^2}\left(\left|\xi\right| - \left|\xi_{\rm eff}\right|\right)^2\tanh^2\left(\sqrt{\frac{C_{\mathcal{B}}}{C_{\mathcal{E}}}}\,\frac{\pi}{\left|\xi_{\rm eff}\right|}\right) & \,, \\
\left|\mathcal{G}\right|_{\rm eq} = \frac{\sqrt{C_{\mathcal{B}}/C_{\mathcal{E}}}}{a_{\rm SM}^2\left|\xi_{\rm eff}\right|^{\phantom{2}}}\left(\left|\xi\right| - \left|\xi_{\rm eff}\right|\right)^2\tanh^2\left(\sqrt{\frac{C_{\mathcal{B}}}{C_{\mathcal{E}}}}\,\frac{\pi}{\left|\xi_{\rm eff}\right|}\right) & \,.
\end{align}
By comparing these expressions with the expressions in Eq.~\eqref{eq:EBGeq}, we obtain a single consistency condition for $\left|\xi_{\rm eff}\right|$, 
\begin{equation}
\label{eq:EQcond}
C_{\mathcal{E}}\,\frac{e^{2\pi\left|\xi_{\rm eff}\right|}}{\left|\xi_{\rm eff}\right|^3} = \frac{1}{a_{\rm SM}^2}\left(\left|\xi\right| - \left|\xi_{\rm eff}\right|\right)^2\tanh^2\left(\sqrt{\frac{C_{\mathcal{B}}}{C_{\mathcal{E}}}}\,\frac{\pi}{\left|\xi_{\rm eff}\right|}\right) \,.
\end{equation}
Therefore, to evaluate the equilibrium estimate, we need to numerically solve this condition for $\left|\xi_{\rm eff}\right|$, for fixed values of $\xi$ and $H$, while making sure that the relations in Eqs.~\eqref{running} and \eqref{eq:muscale} are self-consistently satisfied. The numerical result for $\left|\xi_{\rm eff}\right|$ that we obtain in this way can then be used in Eq.~\eqref{eq:EBGeq}. This procedure defines our numerical results for $\mathcal{E}_{\rm eq}$, $\mathcal{B}_{\rm eq}$, and $\left|\mathcal{G}\right|_{\rm eq}$, which are shown in the form of purple bands in the first row of Fig.~\ref{fig-compare}.

Similarly as in the case of the maximal estimate, we are again able to describe our numerical results in terms of fit functions. This time, the entire relevant information can be encoded in the fit function for $\left|\xi_{\rm eff}\right|$,
\begin{equation}
\label{eq:xieffeq}
\left|\xi_{\rm eff}\right|_{\rm eq} \simeq a_{\rm eq} \ln\left(\left|\xi\right| + b_{\rm eq}\right) + c_{\rm eq}
\end{equation}
with coefficients
\begin{align}
a_{\rm eq} & \simeq \phantom{-} 0.3679 - 0.0004\,\log_{10} H_{12} \,, \\
b_{\rm eq} & \simeq          -  3.3668 + 0.0099\,\log_{10} H_{12} \,, \\
c_{\rm eq} & \simeq \phantom{-} 3.7012 - 0.0152\,\log_{10} H_{12} \,,
\end{align}
where $H_{12}$ denotes again the Hubble rate in units of $10^{12}\,\textrm{GeV}$ [see Eq.~\eqref{eq:gmaxfit}]. In the third row of  Fig.~\ref{fig-compare}, we compare the approximate results for $\mathcal{E}_{\rm eq}$, $\mathcal{B}_{\rm eq}$, and $\left|\mathcal{G}\right|_{\rm eq}$ based on this fit function to the exact numerical results in the first row. Again, we find excellent agreement, in the regime where the expressions in Eq.~\eqref{eq:EBGeq} are valid, i.e., for $\left|\xi_{\rm eff}\right| \gg 3$.

By construction, the fit function in Eq.~\eqref{eq:xieffeq} already takes into account the running of the gauge coupling constant. It is therefore not necessary to work out an independent fit function for $g'$. This differs from the case of the maximal estimate, where we were able to solve the maximization condition in Eq.~\eqref{eq:Gmax} without specifying the coefficient $a_{\rm SM}$. For completeness, we, however, note that the same strategy that eventually led to $g'_{\rm max}$ in Eq.~\eqref{eq:gmax} can be applied in order to determine the self-consistent solution for the gauge coupling constant in the case of the equilibrium estimate,
\begin{equation}
\label{eq:geq}
g_{\rm eq}' = \frac{1}{\sqrt{b\,W_0\left(z_{\rm eq}\right)}} \,, \qquad z_{\rm eq} = \frac{1}{b}\,\exp\left(\frac{1}{b\,g'^2\left(m_Z\right)}\right)\left(\frac{m_Z}{\bar{\mu}_{\rm eq}}\right)^{4/3} \,,
\end{equation}
where the rescaled renormalization scale $\bar{\mu}_{\rm eq}$ is now given by
\begin{equation}
\bar{\mu}_{\rm eq} = \frac{H}{2^{1/4}}\left(\overline{\mathcal{E}}_{\rm eq} + \overline{\mathcal{B}}_{\rm eq}\right)^{1/4} \,,
\end{equation}
\begin{equation}
\overline{\mathcal{E}}_{\rm eq} = \frac{1}{\bar{a}_{\rm SM}^2}\left(\left|\xi\right| - \left|\xi_{\rm eff}\right|\right)^2\tanh^2\left(\sqrt{\frac{C_{\mathcal{B}}}{C_{\mathcal{E}}}}\,\frac{\pi}{\left|\xi_{\rm eff}\right|}\right)  \,,
\end{equation}
\begin{equation}
\overline{\mathcal{B}}_{\rm eq} = \frac{C_{\mathcal{B}}/C_{\mathcal{E}}}{\bar{a}_{\rm SM}^2\left|\xi_{\rm eff}\right|^2}\left(\left|\xi\right| - \left|\xi_{\rm eff}\right|\right)^2\tanh^2\left(\sqrt{\frac{C_{\mathcal{B}}}{C_{\mathcal{E}}}}\,\frac{\pi}{\left|\xi_{\rm eff}\right|}\right) \,.
\end{equation}
Equation~\eqref{eq:geq}, together with $\left|\xi_{\rm eff}\right|_{\rm eq}$ in Eq.~\eqref{eq:xieffeq}, results in excellent agreement (at the level of $10^{-3}\%$) with our numerical results for $g'_{\rm eq}$. Alternatively, we can simply solve the consistency condition in Eq.~\eqref{eq:EQcond} for $g'$,
\begin{equation}
\label{eq:geqex}
g'_{\rm eq} = \left(\frac{1}{C_{\mathcal{E}}}\frac{\left|\xi_{\rm eff}\right|^3}{e^{2\pi \left|\xi_{\rm eff}\right|}}\right)^{1/6}\left[\frac{144\pi^2}{41}\left(\left|\xi\right| - \left|\xi_{\rm eff}\right|\right)\tanh\left(\sqrt{\frac{C_{\mathcal{B}}}{C_{\mathcal{E}}}}\,\frac{\pi}{\left|\xi_{\rm eff}\right|}\right)\right]^{1/3} \,.
\end{equation}
This is an exact expression for $g'_{\rm eq}$, which, however, is more sensitive to deviations of the fit function in Eq.~\eqref{eq:xieffeq} from the exact numerical result. The combination of Eqs.~\eqref{eq:xieffeq} and \eqref{eq:geqex} still results in a good approximation of the exact numerical result for $g'_{\rm eq}$, with the numerical deviations mostly remaining below the percent level. Finally, we are also able to describe the exact numerical result for $g'_{\rm eq}$ with the simple fit function [see also Eq.~\eqref{eq:gmaxfit}]
\begin{equation}
g_{\rm eq}' \simeq 0.4131 + 0.0067 \log_{10} H_{12} + 0.0025 \ln\left(\frac{\left|\xi\right|}{10}\right) \,,
\end{equation}
which reproduces the exact result up to deviations at the level of around $0.1\%$. 

\subsection{Semianalytical fit functions}
\label{subsec:fits}

In the previous two sections, we constructed novel fit functions for the two estimates of the efficiency of gauge-field production that had originally been proposed in  Refs.~\cite{Domcke:2018,Domcke:2019bar}. Now, we turn to our own numerical results based on the gradient-expansion formalism, i.e., the colorful bands for $\Delta = 10^{-6}$, $10^{-4}$, $10^{-2}$, and $1$ in the first row of Fig.~\ref{fig-compare}. To fit our numerical results, we make an ansatz for $X = \mathcal{E}$, $\mathcal{B}$, and $\left|\mathcal{G}\right|$ of the form
\begin{equation}
X\left(\xi,H,\Delta\right) = S_X\left(\xi,H,\Delta\right)\,X_{\Delta = 1}\left(\xi,H\right) \,,
\end{equation}
where $X_{\Delta = 1}$ is supposed to describe our data for $\Delta = 1$ and the function $S_X$ accounts for the suppression of the quantity $X$ if the parameter $\Delta$ is smaller than unity. In fact, it turns out convenient to write $S_X$ as a power of $\Delta$,
\begin{equation}
\label{eq:SXPX}
S_X\left(\xi,H,\Delta\right) = \Delta^{1/P_X} \,, \qquad P_X = P_X\left(\xi,H,\Delta\right) \,.
\end{equation}
We furthermore approximate $X_{\Delta = 1}$ by two different expressions at small and large values of $\left|\xi\right|$,
\begin{equation}
X_{\Delta = 1}\left(\xi,H\right) =
\begin{cases}
X_{\Delta = 1}^<\left(\xi,H\right) \,; & \quad \left|\xi\right| \lesssim \left|\xi\right|_X  \\
X_{\Delta = 1}^>\left(\xi,H\right) \,; & \quad \left|\xi\right| \gtrsim \left|\xi\right|_X  
\end{cases} \,,
\end{equation}
where the threshold value $\left|\xi\right|_X$ at which we switch from one expression to the other is chosen as
\begin{equation}
\left|\xi\right|_{\mathcal{E}} \simeq 4.6 \,, \qquad \left|\xi\right|_{\mathcal{B}} \simeq 5.0 \,, \qquad \left|\xi\right|_{\left|\mathcal{G}\right|} \simeq 4.8 \,, 
\end{equation}
for $\mathcal{E}$, $\mathcal{B}$, and $\left|\mathcal{G}\right|$, respectively. At small $\left|\xi\right|$, we relate our results to the free solution without any backreaction or fermion production, while at large $\left|\xi\right|$, we express our results in relation to the maximal estimate defined in Sec.~\ref{subsec:maxestimate},
\begin{equation}
\label{eq:TXUX}
X_{\Delta = 1}^<\left(\xi,H\right) = T_X\left(\xi,H\right) X_{\rm free}\left(\xi,H\right) \,, \qquad X_{\Delta = 1}^>\left(\xi,H\right) = U_X\left(\xi,H\right) X_{\rm max}\left(\xi,H\right) \,.
\end{equation}
Here, $X_{\rm free}$ ($X = \mathcal{E},\mathcal{B},\left|\mathcal{G}\right|$) is given by the three integral expressions in Eqs.~\eqref{eq:Eintegral}, \eqref{eq:Bintegral}, and \eqref{eq:Gintegral} for $n=0$ and after undoing the replacement $\left|\xi\right| \rightarrow \left|\xi_{\rm eff}\right|$; $X_{\rm max}$ ($X = \mathcal{E},\mathcal{B},\left|\mathcal{G}\right|$) corresponds to our three fit functions for the maximal estimate in Eqs.~\eqref{eq:Emax}, \eqref{eq:Bmax}, and \eqref{eq:Gmx} in combination with our result for $g'_{\rm max}$ in Eq.~\eqref{eq:gmax}. 

The nontrivial information contained in our numerical results is thus captured by the three functions $P_X$, $T_X$, and $U_X$, for each of the three quantities $\mathcal{E}$, $\mathcal{B}$, and $\left|\mathcal{G}\right|$, in Eqs.~\eqref{eq:SXPX} and \eqref{eq:TXUX}. For each of these functions, we make a particular (purely phenomenological) ansatz that turns out to describe our numerical data to very good accuracy,
\begin{align}
\label{eq:PX}
P_X & = 1 + \frac{\exp\left(a_{S_X} + b_{S_X}|\xi| + c_{S_X} |\xi|^2\right)}{|\xi|^{d_{S_X}}} \,,\\
\label{eq:TX}
T_X & = \left[1 + \frac{\exp\left(a_{T_X} + b_{T_X} |\xi|\right)}{1 + |\xi|^{c_{T_X}}}\right]^{-1} \,,\\
\label{eq:UX}
U_X & = a_{U_X} \left[1 - \frac{\exp\left(b_{U_X} + c_{U_X}|\xi|\right)}{|\xi|^{d_{U_X}}}\right] \,.
\end{align}
For each $X$, we hence need to determine 11 fit coefficients: $a_{S_X}$, $b_{S_X}$, $c_{S_X}$, $d_{S_X}$, $a_{T_X}$, $b_{T_X}$, $c_{T_X}$, $a_{U_X}$, $b_{U_X}$, $c_{U_X}$, and $d_{U_X}$. Our results for these, in total, 33 coefficients, which depend on the logarithm of $H$ as well as partially on the logarithm of $\Delta$, are listed in the Appendix. In rows 4 to 7 of Fig.~\ref{fig-compare}, we compare the resulting fit functions for $\mathcal{E}$, $\mathcal{B}$, and $\left|\mathcal{G}\right|$ with the exact numerical results shown in row 1. As before, we find excellent agreement. On a logarithmic scale, our fit functions begin to deviate from the exact numerical results typically only in the third significant digit. In summary, we therefore conclude that the semianalytical fit functions constructed in Secs.~\ref{subsec:maxestimate} and \ref{subsec:eqestimate} as well as in the present section are capable of reproducing all relevant numerical results with very good accuracy. In future work, it will no longer be necessary to repeat the numerical analysis that originally led to these fit functions. 

\subsection{Validation in a specific model}
\label{subsec:validation}

In this section, we test the accuracy of our model-independent approach by comparing it to the exact results for a specific inflationary model. For this purpose, we consider the simple model with a quadratic inflaton potential,
\begin{equation}
    \label{inflaton-potential}
    V(\phi)=\frac{m^{2}\phi^{2}}{2}.
\end{equation}
This quadratic dependence is universal for a wide class of inflaton potentials close to their minima; therefore, validating our formalism in this model will allow us to make more general conclusions. Indeed, since the most efficient gauge-field production occurs close to the end of inflation, the behavior of the inflaton potential far from its minimum is not of great importance for hypermagnetogenesis. In our numerical analysis, we will set $m=6\times 10^{-6}\,M_{\mathrm{P}}$, for concreteness, which is the same value that we used in Ref.~\cite{Gorbar:2021}. However, because of the one-to-one relation between $m$ and the Hubble rate [see Eq.~\eqref{H-slow-roll}], we stress that other values of the inflaton mass will only lead to logarithmic corrections to our results. We therefore expect that the following analysis applies, in fact, to a large range of $m$ values. 

We take the axial-coupling function in a linear form,
\begin{equation}
    \label{axial-coupling}
    I(\phi)=\beta\frac{\phi}{M_{\mathrm{P}}} \,,
\end{equation}
with a dimensionless coupling parameter $\beta$. To obtain the exact numerical results for the generated gauge fields in this model for a given value of $\beta$, we apply the gradient-expansion formalism developed by us in Ref.~\cite{Gorbar:2021}. It is worth noting that the gradient-expansion formalism is itself an approximate method; however, it was shown in Ref.~\cite{Gorbar:2021} that its error compared to the exact mode-by-mode solution can be made as small as $1\%-2\%$ during the whole stage of inflation. As we will see, such an accuracy is much better than that of the model-independent approach; therefore, we can use the gradient-expansion result as a reference solution.

In practice, we proceed as follows. For a given value of $\beta$, we first apply the gradient-expansion formalism of Ref.~\cite{Gorbar:2021} and obtain the hyperelectric and hypermagnetic energy densities, $\rho_E$ and $\rho_B$, as well as the Chern-Pontryagin density $|\langle\bm{E}\cdot\bm{B}\rangle|$ as functions of the number of $e$-folds $N_e$ until the end of inflation. In addition, we also compute the corresponding values of the gauge-field production parameter $\xi$, the Hubble rate $H$, and the damping factor $\Delta$. Knowing the latter three parameters at the same moment of time then allows us to apply our model-independent approach and find the predictions for the generated gauge fields at this moment. Comparing these predictions with the reference solutions, we are able to draw conclusions concerning the accuracy of the model-independent approach.

One may argue that such a usage of the model-independent approach has no advantage compared to the full gradient-expansion formalism because we have to launch the latter method in any case in order to obtain the set of parameters $(\xi,\,H,\,\Delta)$ for the former method. However, this is done only for the purpose of comparing the two methods. Normally, to arrive at model-independent predictions, it suffices to determine the parameters $(\xi,\,H,\,\Delta)$ from some other, much simpler consideration. For instance, the values of $\xi$ and $H$ can be estimated by considering the inflaton dynamics neglecting the backreaction of the generated gauge fields (it is shown in Ref.~\cite{Gorbar:2021} that this is a reasonable approximation for a wide range of parameters in the presence of the Schwinger effect). However, the parameter $\Delta$ cannot be determined by a simple method. Therefore, it is interesting to check whether one can simply use the fixed value $\Delta=1$ in the model-independent approach. This value is well motivated by the following arguments. For small $\xi$, when the dependence of the generated gauge fields on $\Delta$ is strong (see Fig.~\ref{fig-compare}), the gauge fields are rather weak; therefore, the Schwinger conductivity is small compared to the Hubble parameter, and $\Delta$ is indeed close to unity (unless the system had a nontrivial prehistory including a period with very high conductivity). In the opposite case of large $\xi$, the parameter $\Delta$ can be much less than unity; however, the generated fields exhibit only a very weak dependence on $\Delta$, and their values do not differ much from those with $\Delta=1$; see Fig.~\ref{fig-compare}. Therefore, to check the validity of this approximation, we will in addition also apply the model-independent approach to the same values of $(\xi,\,H)$ as before in combination with a fixed value of $\Delta=1$.

\begin{figure}[ht!]
	\centering
	\includegraphics[width=0.97\textwidth]{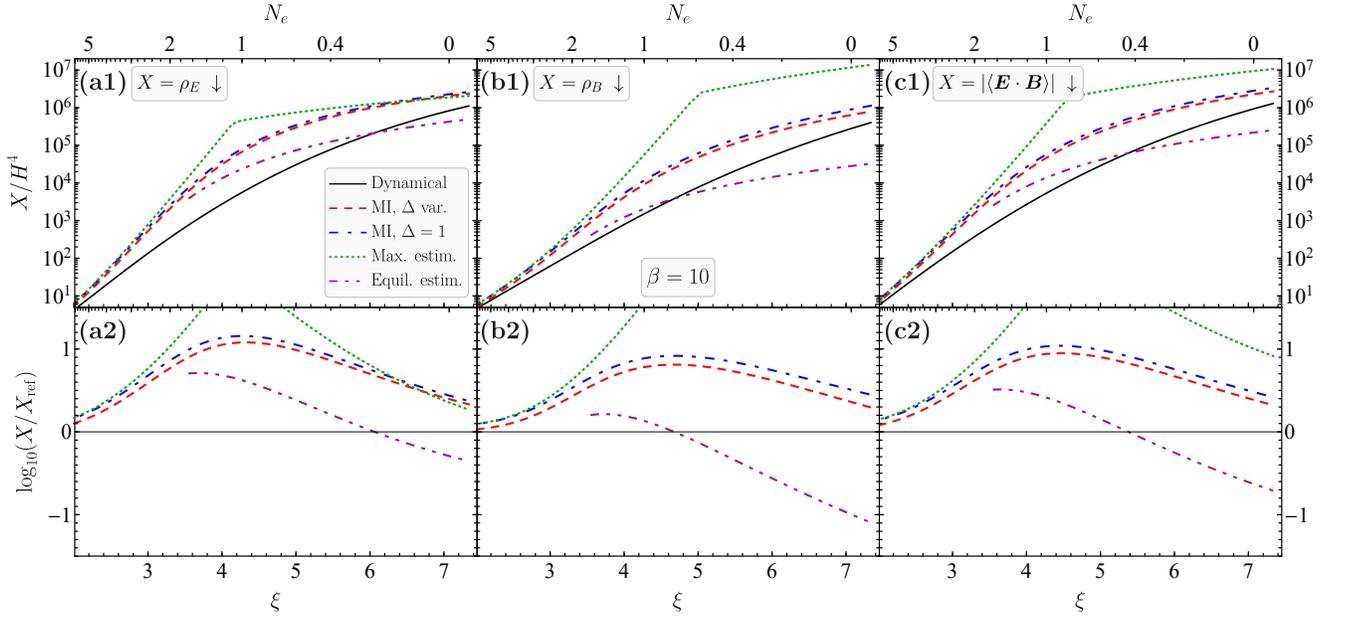}
	\caption{Model-independent (MI) estimates of the generated gauge fields during axion inflation in a specific inflationary model with potential (\ref{inflaton-potential}) and the axial-coupling function (\ref{axial-coupling}) for the coupling parameter $\beta=10$, compared to the numerical results in the same model obtained by means of a full-fledged numerical analysis in the gradient-expansion formalism. The magnitude of the generated hyperelectric energy density $\rho_E$ (column a), hypermagnetic energy density $\rho_B$ (column b), and hypercharge Chern-Pontryagin density $\left|\langle\bm{E}\cdot\bm{B}\rangle\right|$ (column c) are shown in row 1 as functions of the gauge-field production parameter $\xi$ (lower labels on the horizontal axis) and the number of $e$-folds until the end of inflation (upper labels on the horizontal axis). The black solid lines show the numerical results obtained from the full gradient-expansion formalism of Ref.~\cite{Gorbar:2021} for the specific model under consideration. The values of the parameters $\xi$, $H$, and $\Delta$ obtained from this numerical analysis are then used to compute the model-independent predictions for the generated fields: the red dashed lines show the model-independent gradient-expansion predictions for given $\xi$, $H$, and $\Delta$; the blue dashed-dotted lines show the model-independent gradient-expansion results for the given values of $\xi$ and $H$, and a fixed value of $\Delta=1$; the green dotted lines give the maximal estimates; and the purple dashed-dotted lines with double dot show the equilibrium estimates for given $\xi$ and $H$. Row 2 shows the accuracy of the model-independent results compared to the exact numerical solution; the types of curves correspond to those shown in row 1.
	\label{fig-compare10}}
\end{figure}

\begin{figure}[ht!]
	\centering
	\includegraphics[width=0.97\textwidth]{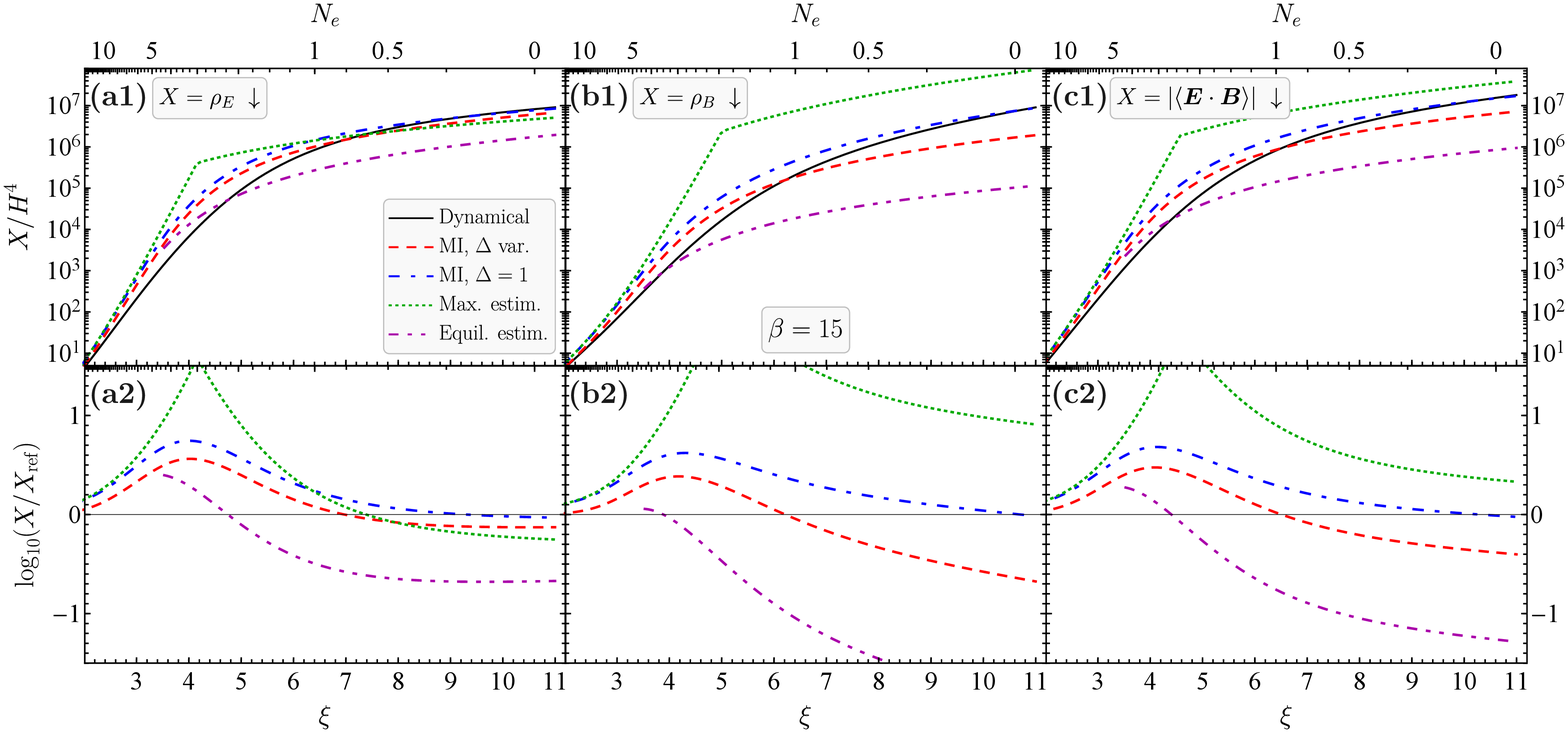}
	\caption{Same quantities as shown in Fig.~\ref{fig-compare10} for the case of the axial-coupling function (\ref{axial-coupling}) with $\beta=15$.
	\label{fig-compare15}}
\end{figure}

\begin{figure}[ht!]
	\centering
	\includegraphics[width=0.97\textwidth]{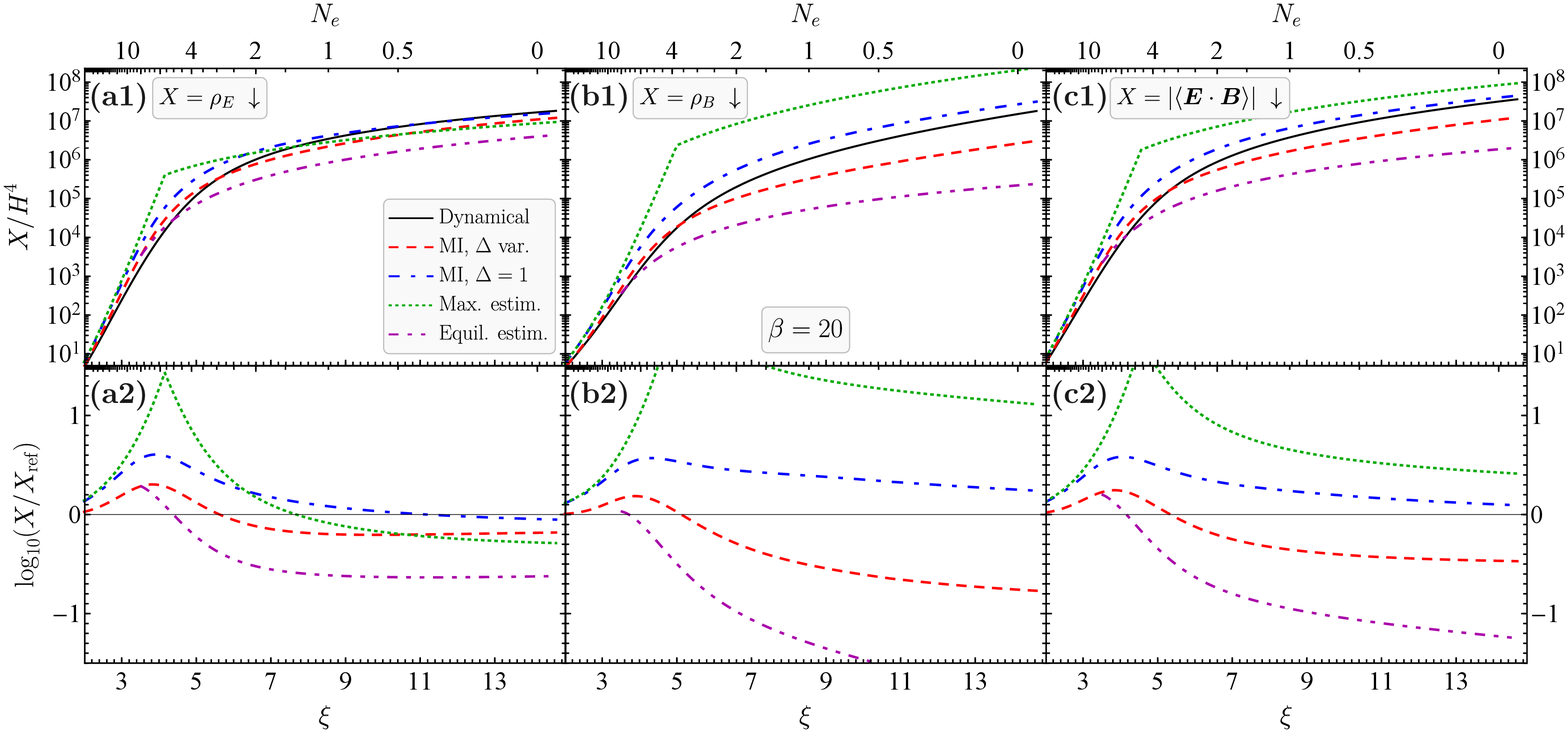}
	\caption{Same quantities as shown in Fig.~\ref{fig-compare10} for the case of the axial-coupling function (\ref{axial-coupling}) with $\beta=20$.
	\label{fig-compare20}}
\end{figure}

Finally, for given $\xi$ and $H$, we also compute the maximal and equilibrium estimates discussed in the previous sections. Thus, for a fixed value of $\beta$, we obtain exact numerical results for the generated gauge fields and four different approximate results. We compare them in Figs.~\ref{fig-compare10}, \ref{fig-compare15}, and \ref{fig-compare20} for $\beta=10$, 15, and 20, respectively. The upper rows of the respective figures show the magnitude of the hyperelectric energy density $\rho_{E}$ (column~a), hypermagnetic energy density $\rho_B$ (column b), and hypercharge Chern-Pontryagin density $\left|\langle\bm{E}\cdot\bm{B}\rangle\right|$ (column c) as functions of the parameter $\xi$. Here, note that for a fixed $\beta$ there is a one-to-one correspondence between the number of $e$-folds before the end of inflation and the value of the parameter $\xi$; we show the corresponding values of $N_{e}$ on the top horizontal axes. The black solid curves show the results of the full gradient-expansion formalism (the reference solution), while the curves of different colors and dashing types show the approximate solutions: the model-independent predictions for given values of $\xi$, $H$, and $\Delta$ (red dashed lines); the model-independent results for given $\xi$, $H$, and fixed $\Delta=1$ (blue dashed-dotted lines); the maximal estimates for given $\xi$ and $H$ (green dotted lines); and the equilibrium estimates for given $\xi$ and $H$ (purple dashed-dotted lines with double dot). The lower rows of the respective figures represent the deviation between the approximate solutions and the reference one on a logarithmic scale.

\begin{figure}[ht!]
	\centering
	\includegraphics[width=0.97\textwidth]{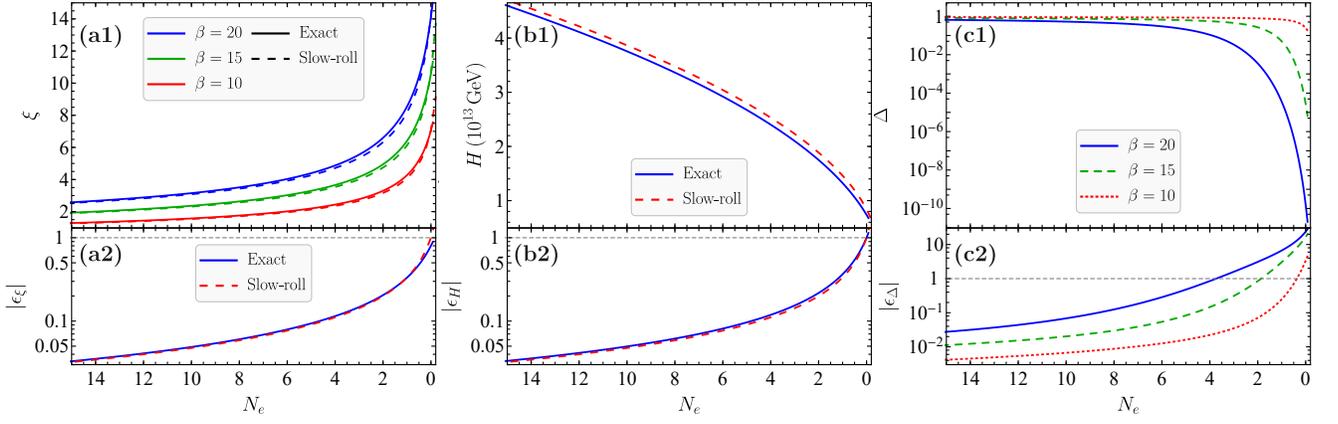}
	\caption{Row 1: gauge-field production parameter $\xi$ (column~a), Hubble parameter $H$ (column~b), and damping parameter $\Delta$ (column~c) as functions of the number of $e$-folds to the end of inflation $N_{e}$ in the inflationary model with the quadratic potential (\ref{inflaton-potential}) with $m=6\times 10^{-6}\,M_{\mathrm{P}}$ and the axial-coupling function (\ref{axial-coupling}) for three values of the coupling parameter $\beta=10$, $15$, and $20$. Row 2: adiabaticity parameters $\epsilon_{P}=(1/H)(\dot{P}/P)$ for the corresponding quantities. In panels~a1, a2, b1, and b2, the solid lines show exact numerical results obtained in the full gradient-expansion formalism, while the dashed lines represent the approximate slow-roll expressions (\ref{H-slow-roll})--(\ref{eps-xi-slow-roll}). In panels~c1 and c2, only exact numerical results are shown; the three types of curves correspond to the three values of $\beta$.
	\label{fig-epsilon}}
\end{figure}

Let us now analyze and comment on the obtained numerical results. First of all, we mention that the model-independent approach presented in this paper allows us to estimate the magnitude of the generated gauge field within at most 1 order of magnitude in comparison to the exact numerical results. The maximal deviation occurs very close to the end of inflation (during the last one to two $e$-folds), where the change in the parameters $\xi$, $H$, and $\Delta$ cannot be considered to be adiabatically slow (see Fig.~\ref{fig-epsilon} and its discussion below). This deviation can be easily understood since the model-independent approach assumes that the above-mentioned parameters are constant. Second, the model-independent approach with a fixed value of $\Delta=1$ typically results in a slightly worse agreement (although it is still comparable with the model-independent approach with the exact value of $\Delta$). However, for larger values of $\beta$, at the very end of inflation, its predictions are accidentally even in better accordance with the exact result (compare the red and blue curves in Figs.~\ref{fig-compare15} and \ref{fig-compare20}). Third, the model-independent approach typically gives more accurate results than the equilibrium or maximal estimates. Only for the hyperelectric energy density, it sometimes happens that the latter estimates have a comparable accuracy with the model-independent approach.

Finally, let us discuss the reasons for the deviation of the approximate methods considered above from the exact numerical solution. As we already pointed out, all approximate methods rely on the fact that their input parameters $\xi$, $H$, and $\Delta$ are constant. Such an approximation would be reasonable if these parameters were changing adiabatically slowly, i.e., their change during the Hubble time was much smaller than the absolute value of the parameter. This condition can be characterized by the adiabaticity parameter $\epsilon_{P}$ defined for any $P=\{\xi,\,H,\,\Delta\}$ in the following way:
\begin{equation}
    \epsilon_{P}=\frac{1}{H}\frac{\dot{P}}{P}.
\end{equation}
In particular, the parameter $\epsilon_{H}$ is the well-known slow-roll parameter that controls when inflation terminates. Some preliminary estimates for the parameters $\epsilon_{H}$ and $\epsilon_{\xi}$ can be obtained from the slow-roll analysis. For the inflationary model with the scalar potential (\ref{inflaton-potential}) in the absence of any backreaction from the generated gauge fields, it is possible to analytically find the dependence of the inflaton field on the number of $e$-folds before the end of inflation,
\begin{equation}
    \phi(N_{e})=M_{\mathrm{P}}\sqrt{2(1+2N_{e})},
\end{equation}
from which one immediately obtains the slow-roll expressions for $\epsilon_{H}$, $\epsilon_{\xi}$, given the coupling function $I(\phi)=\beta\phi/M_{\mathrm{P}}$,
\begin{equation}
\label{H-slow-roll}
H\simeq \frac{m\phi}{\sqrt{6}M_{\mathrm{P}}}=\frac{m}{\sqrt{3}}\sqrt{1+2N_{e}},
\end{equation}
\begin{equation}
\label{xi-slow-roll}
|\xi|\simeq \frac{\beta M_{\mathrm{P}}}{2}\left|\frac{V'}{V}\right|=\frac{\beta M_{\mathrm{P}}}{\phi}=\frac{\beta}{\sqrt{2(1+2N_{e})}},
\end{equation}
\begin{equation}
\label{eps-H-slow-roll}
    |\epsilon_{H}|\simeq \frac{M_{\mathrm{P}}^{2}}{2}\Big(\frac{V'}{V}\Big)^{2}=\frac{2M_{\mathrm{P}}^{2}}{\phi^{2}}=\frac{1}{1+2N_{e}},
\end{equation}
\begin{equation}
\label{eps-xi-slow-roll}
    |\epsilon_{\xi}|\simeq M_{\mathrm{P}}^{2}\Big|\frac{V''}{V}-\frac{V^{\prime 2}}{V^{2}}\Big|=\frac{2M_{\mathrm{P}}^{2}}{\phi^{2}}=\frac{1}{1+2N_{e}}.
\end{equation}

Concerning the parameter $\epsilon_{\Delta}$, it follows from Eq.~(\ref{eq:delta}) that
\begin{equation}
    \epsilon_{\Delta}=\frac{\sigma}{H}=2s.
\end{equation}
Although we cannot derive any slow-roll estimates for this parameter, it is clear that it increases when the generated gauge field becomes stronger. Figure~\ref{fig-epsilon} shows the parameters $\xi$, $H$, and $\Delta$ during the last 15 $e$-folds of inflation (row~1) and the corresponding parameters $\epsilon_{\xi}$, $\epsilon_{H}$, and $\epsilon_{\Delta}$ (row~2).

In panel~a1 of Fig.~\ref{fig-epsilon}, we show the parameter $\xi$ for $\beta=10$ (red line), $\beta=15$ (green line), and $\beta=20$ (blue line). The solid lines correspond to the exact numerical result, while the dashed lines of the same colors give the slow-roll estimates according to Eq.~(\ref{xi-slow-roll}). Since $\xi\propto \beta$, the parameter $\epsilon_{\xi}$ does not depend on $\beta$. The same also holds for $H$ and $\epsilon_{H}$, if there is no backreaction from the generated gauge fields on the inflaton dynamics (this is indeed the case in our model). Therefore, in panels~a2, b1, and b2, we show exact results for $\epsilon_{\xi}$, $H$, and $\epsilon_{H}$, respectively, by the blue solid lines and the corresponding slow-roll estimates by the red dashed lines. It is worth noting that in the slow-roll approximation $\epsilon_{\xi}=\epsilon_{H}$. As expected, $\epsilon_{H}$ and $\epsilon_{\xi}$ tend to unity when inflation ends. Panel~c1 of Fig.~\ref{fig-epsilon} shows the dependence of the parameter $\Delta$ on the number of $e$-folds for $\beta=10$ (red dotted line), $\beta=15$ (green dashed line), and $\beta=20$ (blue solid line). Far from the end of inflation, this parameter is very close to unity because the gauge fields are weak and, consequently, the Schwinger conductivity is small (see panel~c2, where the corresponding values of $\epsilon_{\Delta}=\sigma/H$ are shown). However, when inflation ends, it becomes exponentially small. The corresponding adiabaticity parameter exceeds unity (this happens earlier for larger $\beta$), and the adiabatic approximation fails. The behavior illustrated in Fig.~\ref{fig-epsilon} thus explains why the approximate model-independent results begin to differ from the exact solution close to the end of inflation.

\section{Conclusion}
\label{sec-concl}

The description of gauge-field production during axion inflation is relevant for a variety of phenomena, ranging from primordial magnetic fields over baryogenesis to primordial gravitational waves and black holes. In the presence of the Schwinger effect, this process becomes highly nonlinear and typical approaches dealing with separate gauge-field modes in Fourier space become inapplicable. To overcome this difficulty, we proposed in Ref.~\cite{Gorbar:2021} a novel gradient-expansion formalism that operates with a set of bilinear gauge-field quantities in coordinate space and that allows us to build a complete and self-consistent system of equations for studying hypermagnetogenesis during axion inflation in the presence of nonlinear effects, such as the backreaction of the generated fields on the inflaton field and the Schwinger effect. Although this system of equations can be explicitly solved in a given inflationary model, such an analysis is quite complicated and requires some computational efforts. However, for many practical purposes, it would be desirable and convenient to have some model-independent predictions allowing one to estimate the magnitude of the generated gauge fields without carrying out a complicated numerical analysis. In the free case involving no nonlinear effects at all, such model-independent estimates are well known in the literature: the electric energy density, magnetic energy density, and Chern-Pontryagin density, respectively, scale like $H^4\,e^{2\pi\xi}/\xi^3$, $H^4\,e^{2\pi\xi}/\xi^5$, and $H^4\,e^{2\pi\xi}/\xi^4$ for large $\xi$, if all nonlinear effects can be neglected. The aim of the present work was to generalize these model-independent estimates to the full SM case in the presence of Schwinger pair production of all SM fermions. 

The fact that gauge-field production during axion inflation is controlled by $\xi$ and $H$ has been known for a long time. We recall once more that the parameter $\xi$ characterizes the velocity of the inflaton field and $H$ denotes the Hubble expansion rate. Both parameters can be easily determined from the standard slow-roll analysis in a concrete inflationary model. In the presence of Schwinger pair production, there appears one additional parameter $\Delta$, which describes the damping of the vacuum gauge-field fluctuations due to the finite conductivity of the Universe on subhorizon scales~\cite{Gorbar:2021}. This parameter depends on the prehistory of the system [see Eq.~(\ref{eq:delta})] and thus makes its evolution nonlocal in time, which complicates the model-independent analysis of gauge-field production to some degree. 

The main assumption behind our analysis was that gauge-field production at some moment of time during axion inflation is determined by the momentary values of the parameters $\xi$, $H$, and $\Delta$. This assumption is based on the fact that the exponentially fast cosmic expansion quickly dilutes the gauge fields generated at earlier times and thus the dominating contribution to the magnitude of the gauge fields at some moment of time originates from the modes that crossed the horizon just shortly before. A convenient criterion for the validity of this assumption is the adiabaticity of evolution of all three parameters: their change during the Hubble time must be much less than the absolute value of the parameter itself. For the parameters $\xi$ and $H$, this adiabatically slow evolution indeed holds during slow-roll inflation. As for the parameter $\Delta$, it changes slowly only in the weak-field regime when the corresponding Schwinger conductivity is much less than the Hubble parameter, $\sigma\ll H$; see Fig.~\ref{fig-epsilon} for an explicit example.

To derive our new model-independent results, we employed the gradient-expansion formalism presented in Ref.~\cite{Gorbar:2021}, for which we used constant values of $\xi$, $H$, and $\Delta$ as input parameters. The main difference compared to our earlier model-dependent analysis in Ref.~\cite{Gorbar:2021} is that, in the full system of equations, these parameters are computed self-consistently by considering the evolution of the inflaton field and scale factor together with the gauge fields. However, as we have shown in this paper, the approximation of constant $\xi$, $H$, and $\Delta$ indeed works well when their respective time variation in a concrete model remains adiabatically slow. In our numerical analysis, we fixed the parameters $\xi$, $H$, and $\Delta$ and looked for the stationary solution of our system of equations, which provided us with the prediction for the generated gauge field. Then, we scanned over wide ranges of parameter values that are sufficient for most physical applications, namely, $1<\xi<15$, $10^{8}\,\text{GeV}<H<10^{14}\,\text{GeV}$, and $10^{-6}<\Delta<1$. Our numerical results for the hyperelectric and hypermagnetic energy densities, $\rho_{E}$ and $\rho_{B}$, as well as the Chern-Pontryagin density $|\langle\bm{E}\cdot\bm{B}\rangle|$ are summarized in Fig.~\ref{fig-compare}, which is the main result of our study. There, we also compare our predictions to estimates that had previously been obtained in the literature, in particular, the maximal and equilibrium estimates derived in Refs.~\cite{Domcke:2018,Domcke:2019bar}. They were derived as upper and lower constraints on $|\langle\bm{E}\cdot\bm{B}\rangle|$ without taking into account the damping of vacuum fluctuations by the parameter $\Delta$, i.e., for $\Delta=1$. Our predictions for $|\langle\bm{E}\cdot\bm{B}\rangle|$ in the case $\Delta=1$ lie between the two above-mentioned estimates, thus being in good accordance with them.

For small values of $\Delta$, the results change dramatically only for small values of $\xi$, when the gauge fields are weak. Indeed, in such a case, each new mode crossing the horizon makes an important contribution to the total energy density. Damping of these new modes thus significantly changes the resulting gauge field. On the contrary, for the case of strong gauge fields, the contributions of new modes crossing the horizon are small compared to those that are already outside the horizon and are enhanced due to the axion coupling. Therefore, damping of these new modes by the $\Delta$ parameter makes a small impact on the generated field. These features are clearly seen from Fig.~\ref{fig-compare}.

By construction, our new model-independent estimates do not account for possible backreaction effects, as they are based on the assumption of adiabatically slowly varying values of $\xi$ and $H$. This, however, does not limit the range of applicability of our results as severely as one may naively think. In Ref.~\cite{Gorbar:2021}, we showed that Schwinger pair production often suppresses backreaction effects in scenarios in which it would otherwise be relevant. In the presence of the Schwinger effect, backreaction therefore only occurs in extreme regions of parameter space. As long as it is negligible, our results are applicable and can be considered as the straightforward generalization of the corresponding expressions in the free case (i.e., the expressions proportional to $H^4\,e^{2\pi\xi}/\xi^3$, $H^4\,e^{2\pi\xi}/\xi^5$, and $H^4\,e^{2\pi\xi}/\xi^4$). To make our results more accessible and easier to use, we provide semianalytical fit functions that describe our entire numerical data with very high accuracy across the full range of parameter values that we considered in this paper.

To validate our model-independent results, we considered a concrete inflationary model with potential $V(\phi)=m^{2}\phi^{2}/2$. Although such a potential is already discarded by cosmic microwave background observations, it is still worth considering because many other inflaton potentials can be approximated by $m^{2}\phi^{2}/2$ close to their minima, and this region appears to be the most important for the generation of gauge fields, which occurs during the last few $e$-folds of inflation (this is because the generation is determined by the parameter $\xi \propto \dot{\phi}$ and the inflaton velocity $\dot{\phi}$ typically is the largest close to the end of inflation). We implemented the inflationary model with potential $V(\phi)=m^{2}\phi^{2}/2$ in the full gradient-expansion formalism. Then, we used the exact values of the parameters $\xi$, $H$, and $\Delta$ at a sequence of moments of time close to the end of inflation and launched the model-independent approach for these values of parameters (again treating them as constants). By comparing these approximate results with the results of the self-consistent gradient-expansion formalism, we conclude that the model-independent results indeed can be used to estimate the magnitude of the produced gauge fields with an error typically less than 1 order of magnitude. This is a significant improvement over previous estimates, which spanned several orders of magnitude. The main accomplishment of the present paper is therefore a significant reduction in the theoretical error in the description of hypermagnetogenesis during axion inflation. The largest error is reached close to the end of inflation, where the adiabaticity conditions for $\xi$, $H$, and $\Delta$ break down, while far from the end of inflation, the accuracy of the model-independent result is much better. Moreover, we show that one can even use the fixed value $\Delta=1$ during the whole stage of inflation and the accuracy of our model-independent predictions remains of the same order, which facilitates the usage of our results.

We stress again that our model-independent results are particularly well suited to estimate the efficiency of gauge-field production during inflation. Toward the end of inflation, where the slow-roll approximation breaks down, only estimates within 1 order of magnitude are possible (which, however, still improves on earlier estimates).
In future work, we will therefore turn to a dedicated and more precise description of the initial conditions for reheating after the end of inflation. This analysis will then allow one to connect the dynamics of hypermagnetogenesis during axion inflation to the subsequent evolution during reheating and the radiation-dominated stage after axion inflation. 

\vspace{-3mm}

\begin{acknowledgments}	
We thank Valerie Domcke and Kyohei Mukaida for helpful discussions on the lower integration boundary in Eq.~\eqref{eq:delta}. O.~O.~S.\ is grateful to the CERN Theory Group, where part of this work was done, for its kind hospitality. The work of E.~V.~G. was supported by the National Research Foundation of Ukraine Project No.~2020.02/0062. The work of K.~S. was supported by the European Union's Horizon 2020 Research and Innovation Programme under Grant No.~796961, ``AxiBAU''. The work of O.~O.~S. was supported by the ERC-AdG-2015 Grant No.~694896 and the Swiss National Science Foundation Excellence Grant No.~200020B\_182864. The work of S.~I.~V. was supported by the Swiss National Science Foundation Grant No.~SCOPE IZSEZ0 206908.	
\end{acknowledgments}

\appendix

\section{Numerical fit coefficients}

We list our results for the numerical fit coefficients appearing in Eqs.~\eqref{eq:PX}, \eqref{eq:TX}, and \eqref{eq:UX}. For $X = \mathcal{E}$,
\begin{align}
a_{S_{\mathcal{E}}} & \simeq - 13.6184 + 0.1408\,\log_{10} H_{12} - 0.0842\,\log_{10} \Delta \,,\\
b_{S_{\mathcal{E}}} & \simeq -  4.9636 + 0.0299\,\log_{10} H_{12} - 0.6690\,\log_{10} \Delta \,,\\
c_{S_{\mathcal{E}}} & \simeq    0.1750 - 0.0006\,\log_{10} H_{12} + 0.0263\,\log_{10} \Delta \,,\\
d_{S_{\mathcal{E}}} & \simeq - 21.8036 + 0.1577\,\log_{10} H_{12} - 1.8936\,\log_{10} \Delta \,,\\
a_{T_{\mathcal{E}}} & \simeq -  8.6139 + 0.0523\,\log_{10} H_{12} \,,\\
b_{T_{\mathcal{E}}} & \simeq    4.7885 + 0.0402\,\log_{10} H_{12} \,,\\
c_{T_{\mathcal{E}}} & \simeq    6.6715 + 0.1143\,\log_{10} H_{12} \,,\\
a_{U_{\mathcal{E}}} & \simeq    1.6893 + 0.0004\,\log_{10} H_{12} \,,\\
b_{U_{\mathcal{E}}} & \simeq -  0.0412 + 0.0633\,\log_{10} H_{12} \,,\\
c_{U_{\mathcal{E}}} & \simeq -  0.8612 + 0.0096\,\log_{10} H_{12} \,,\\
d_{U_{\mathcal{E}}} & \simeq -  2.5081 + 0.0751\,\log_{10} H_{12} \,.
\end{align}
For $X = \mathcal{B}$,
\begin{align}
a_{S_{\mathcal{B}}} & \simeq - 11.7274 + 0.1166\,\log_{10} H_{12} + 0.6398\,\log_{10} \Delta \,,\\
b_{S_{\mathcal{B}}} & \simeq - 3.1377  + 0.0162\,\log_{10} H_{12} - 0.1007\,\log_{10} \Delta \,,\\
c_{S_{\mathcal{B}}} & \simeq   0.0829  - 0.0003\,\log_{10} H_{12} + 0.0044\,\log_{10} \Delta \,,\\
d_{S_{\mathcal{B}}} & \simeq - 15.8943 + 0.1065\,\log_{10} H_{12} + 0.0206\,\log_{10} \Delta \,,\\
a_{T_{\mathcal{B}}} & \simeq -  9.4153 + 0.1065\,\log_{10} H_{12} \,,\\
b_{T_{\mathcal{B}}} & \simeq    3.8583 + 0.0576\,\log_{10} H_{12} \,,\\
c_{T_{\mathcal{B}}} & \simeq    3.9415 + 0.2121\,\log_{10} H_{12} \,,\\
a_{U_{\mathcal{B}}} & \simeq    0.1217 + 0.0014\,\log_{10} H_{12} \,,\\
b_{U_{\mathcal{B}}} & \simeq -  2.7356 - 0.0477\,\log_{10} H_{12} \,,\\
c_{U_{\mathcal{B}}} & \simeq -  1.2804 - 0.0200\,\log_{10} H_{12} \,,\\
d_{U_{\mathcal{B}}} & \simeq -  5.5354 - 0.0867\,\log_{10} H_{12} \,.
\end{align}
For $X = \left|\mathcal{G}\right|$,
\begin{align}
a_{S_{\left|\mathcal{G}\right|}} & \simeq - 12.2423 + 0.1270\,\log_{10} H_{12} + 0.4276\,\log_{10} \Delta \,,\\
b_{S_{\left|\mathcal{G}\right|}} & \simeq -  3.0879 + 0.0230\,\log_{10} H_{12} - 0.1879\,\log_{10} \Delta \,,\\
c_{S_{\left|\mathcal{G}\right|}} & \simeq    0.0798 - 0.0005\,\log_{10} H_{12} + 0.0066\,\log_{10} \Delta \,,\\
d_{S_{\left|\mathcal{G}\right|}} & \simeq - 16.3383 + 0.1292\,\log_{10} H_{12} - 0.3635\,\log_{10} \Delta \,,\\
a_{T_{\left|\mathcal{G}\right|}} & \simeq -  8.0575 + 0.0274\,\log_{10} H_{12} \,,\\
b_{T_{\left|\mathcal{G}\right|}} & \simeq    4.8661 + 0.0167\,\log_{10} H_{12} \,,\\
c_{T_{\left|\mathcal{G}\right|}} & \simeq    7.5636 + 0.0318\,\log_{10} H_{12} \,,\\
a_{U_{\left|\mathcal{G}\right|}} & \simeq    0.4511 + 0.0027\,\log_{10} H_{12} \,,\\
b_{U_{\left|\mathcal{G}\right|}} & \simeq -  1.3945 + 0.0150\,\log_{10} H_{12} \,,\\
c_{U_{\left|\mathcal{G}\right|}} & \simeq -  1.0690 - 0.0045\,\log_{10} H_{12} \,,\\
d_{U_{\left|\mathcal{G}\right|}} & \simeq -  4.0335 + 0.0004\,\log_{10} H_{12} \,.
\end{align}

 \end{document}